\documentclass[twocolumn]{aastex63}
\usepackage{bm}

\newcommand{\mytilde}{\raise.19ex\hbox{$\scriptstyle\sim$}}

\usepackage[utf8]{inputenc}
\usepackage{CJK}
\usepackage{amsmath}
\usepackage{lineno}
\usepackage{tabularx}

\received{}
\revised{}
\accepted{}

\submitjournal{}

\shorttitle{}
\shortauthors{Cha et al. }

\begin{document}
\title{MrMARTIAN: A Multi-resolution Mass Reconstruction Algorithm Combining Free-form and Analytic Components}

\correspondingauthor{M. James Jee}
\email{sang6199@yonsei.ac.kr, mkjee@yonsei.ac.kr}

\author[0000-0001-7148-6915]{Sangjun Cha}
\affiliation{Department of Astronomy, Yonsei University, 50 Yonsei-ro, Seoul 03722, Korea}
\author[0000-0002-5751-3697]{M. James Jee}
\affiliation{Department of Astronomy, Yonsei University, 50 Yonsei-ro, Seoul 03722, Korea}
\affiliation{Department of Physics and Astronomy, University of California, Davis, One Shields Avenue, Davis, CA 95616, USA}

\begin{abstract}
We present {\tt MrMARTIAN} (Multi-resolution MAximum-entropy Reconstruction Technique Integrating Analytic Node), a new hybrid strong lensing (SL) modeling algorithm. By incorporating physically motivated analytic nodes into the free-form method {\tt MARS}, {\tt MrMARTIAN} enables stable and flexible mass reconstructions while mitigating oversmoothing in the inner mass profile. 
Its multi-resolution framework increases the degrees of freedom in regions with denser strong lensing constraints, thereby enhancing computational efficiency for a fixed number of free parameters. 
We evaluate the performance of {\tt MrMARTIAN} using publicly available simulated SL data and find that it consistently outperforms {\tt MARS} in recovering both mass and magnification.
In particular, it delivers significantly more stable reconstructions when multiple images are sparsely distributed.
Finally, we apply {\tt MrMARTIAN} to the galaxy cluster MACS J0416.1--2403, incorporating two analytic nodes centered on the northeastern and southwestern BCGs.
Our mass model, constrained by 412 multiple images, achieves an image-plane rms scatter of $\mytilde0\farcs11$, the smallest to date for this dataset.
\end{abstract}


\section{Introduction}\label{sec:intro}

As the most massive gravitationally bound structures in the universe, galaxy clusters have been the subject of extensive studies. Galaxy clusters enable the study of a wide range of topics, including the evolution of large-scale structures \citep[e.g.,][]{2015Natur.528..105E, 2020MNRAS.494.5473K, 2024NatAs.tmp....9H}, cosmology \citep[e.g.,][]{2017MNRAS.470.1809A, 2018ApJ...865..122M, 2022A&A...657A..83C}, and the  properties of dark matter (DM) \citep[e.g.,][]{2004ApJ...606..819M, 2019MNRAS.488.3646R, 2024A&A...687A.270R}. Among various techniques, strong gravitational lensing (SL) is one of the most powerful methods for studying galaxy clusters. 

Thanks to its unparalleled sensitivity to fine details in mass distributions, SL enables high-resolution and precise mass reconstructions around cluster cores and their inner profiles \citep[e.g.,][]{2013ApJ...762L..30Z, 2016ApJ...819..114K, 2016MNRAS.461.2126S, 2021A&A...645A.140B, 2023ApJ...951..140C}. 
Also, massive SL galaxy clusters  serve as cosmic telescopes, magnifying faint background objects \citep[e.g.,][]{2022Natur.603..815W, 2023ApJ...949L..34H, 2024Natur.628...57F}. 
In addition, SL can be used to infer the Hubble constant $H_0$ by measuring time delays between multiple images of time-varying sources such as quasars or supernovae \citep[e.g.,][]{2015Sci...347.1123K, 2020MNRAS.498.1420W, 2025ApJ...979...13P}.  

A number of SL modeling algorithms exist. Broadly, there are two types of SL modeling methods: parametric and free-form. Parametric methods utilize analytic profiles such as NFW \citep{1996ApJ...462..563N}, dPIE \citep{2007arXiv0710.5636E} or PIEMD \citep{1993ApJ...417..450K}.  Parametric methods reconstruct mass distributions by superposing these analytic profiles, with their positions constrained by the locations of cluster member galaxies. In contrast, free-form methods do not assume specific profiles for mass modeling. The mass is reconstructed on a grid without incorporating light distributions as priors.

Both approaches have advantages and disadvantages. Parametric methods typically require fewer free parameters for modeling and can yield numerically stable results even with a small number of multiple images. 
However, the limited number of degrees of freedom restricts their ability to model complex and unexpected mass distributions. This limitation would be particularly concerning {\it if} the true mass distribution deviates from the light distribution.
On the other hand, free-form methods offer excellent flexibility, but their high dimensionality often leads to degeneracies and unphysical features in the resulting mass models.

Recently, \citet{2022ApJ...931..127C} introduced a new grid-based free-form MAximum-entropy ReconStruction ({\tt MARS}) algorithm, which employs maximum cross-entropy regularization. Thanks to this additional constraint, {\tt MARS} can provide a quasi-unique solution\footnote{Here, ``quasi-unique" indicates that, for a fixed dataset, reconstructions with different initial conditions converge to consistent mass distributions for the region constrained by the SL data. See \citet{2022ApJ...931..127C} for more details.} and suppress spurious fluctuations \citep{2023ApJ...951..140C, 2024ApJ...961..186C, 2025ApJ...987L..15C}. 
However, despite these successes, {\tt MARS} still has limitations. One such limitation is the smoothing of density peaks, which arises from both regularization and the finite grid size, especially in regions where SL constraints are sparse. Another limitation is the model's dependence on priors in regions lacking SL constraints. Since free-form models are solely constrained by the multiple images, an asymetric or sparse distribution of SL constraints can introduce biases in the reconstructed mass distribution.
\citep{2024SSRv..220...19N}.

In this study, we propose a new hybrid Multi-resolution MAximum-entropy Reconstruction Technique Integrating Analytic Node ({\tt MrMARTIAN}) to mitigate the aforementioned drawbacks of {\tt MARS}. 
Although we choose a truncated NFW profile \citep{2009JCAP...01..015B} as an analytic node in this study, {\tt MrMARTIAN} can use a wide range of analytic profiles.
By combining both free-form and parametric approaches, {\tt MrMARTIAN} can leverage the advantages of both methods. 
In addition, we implement a multi-resolution framework in lens modeling. By assigning a higher resolution grid (i.e., more degrees of freedom) to regions with denser SL constraints, the total number of free parameters can be reduced without a significant loss of reconstruction quality. This enhancement enables {\tt MrMARTIAN} to achieve improved computational efficiency. 

To evaluate the performance of the {\tt MrMARTIAN} algorithm, we reconstruct lens models of simulated galaxy clusters. We use the publicly available synthetic clusters Ares and Hera from \citet{2017MNRAS.472.3177M}, which are designed to mimic galaxy clusters observed by the Hubble Frontier Fields (HFF) program \citep{2015ApJ...800...84C, 2017ApJ...837...97L}. 
We also apply {\tt MrMARTIAN} to the real galaxy cluster MACS J0416.1-2403 ($z=0.396$, hereafter MACSJ0416). As one of the targets of the HFF program, MACSJ0416 is well-known for its large number of multiple images \citep{2021A&A...646A..83R, 2023ApJ...951..140C, 2023A&A...674A..79B}. 
Moreover, MACSJ0416 has also been observed by JWST as part of the Prime Extragalactic Areas for Reionization and Lensing Science \citep[PEARLS;][]{2023AJ....165...13W} and the CAnadian NIRISS Unbiased Cluster Survey \citep[CANUCS;][]{2022PASP..134b5002W} programs. Based on the HFF and JWST data, lensing studies of MACSJ0416 have been published by both collaborations \citep{2023A&A...679A..31D, 2024A&A...681A.124D, 2025A&A...696A..15R}. \citet{2025A&A...696A..15R} report 415 multiple images across the MACSJ0416 field of view (FOV), which is the largest SL dataset to date.  
This extensive dataset makes MACSJ0416 an ideal testbed for evaluating whether {\tt MrMARTIAN} is applicable in the current JWST era when hundreds of multiple images become available. 

This paper is organized as follows. In \textsection\ref{sec:data}, we introduce the data used in the lensing analysis. We describe our new hybrid algorithm for mass reconstruction in \textsection\ref{sec:method}. We present and discuss our results of MACSJ0416 in \textsection\ref{sec:result} and \textsection\ref{sec:discuss}, respectively. We conclude in \textsection\ref{sec:conclusion}.
Unless stated otherwise, this paper assumes a flat $\Lambda$CDM cosmology with the dimensionless Hubble parameter $h=0.7$ and the matter density parameter $\Omega_{M}=1-\Omega_{\Lambda}=0.3$. The plate scale at the cluster redshift ($z=0.396$) is $5.34 ~\rm kpc ~\rm arcsec^{-1}$.
We make our lens model publicly available\footnote{\url{https://github.com/sang6199/MACS-J0416.1-2403-result-maps}}.

\section{Data} \label{sec:data}
\subsection{Mock Cluster Data}
\label{mock_data}
We use publicly available SL data for two simulated galaxy clusters, Ares and Hera, from \citet{2017MNRAS.472.3177M}\footnote{\url{http://pico.oabo.inaf.it/~massimo/Public/FF/index.html}}. 
Ares is generated using the semianalytic code {\tt MOKA} \citep{2012MNRAS.421.3343G} and is located at $z=0.5$. Hera, at $z=0.507$, is created from a zoom-in, high-resolution resimulation of a cosmological simulation presented in \citet{2014MNRAS.438..195P}. 
Ares has two major halos: one to the northwest with mass $M\sim8.8\times10^{14} h^{-1}~  M_{\odot}$, and the other to the southeast with mass $M\sim1.32\times10^{15} h^{-1} ~ M_{\odot}$, separated by a projected distance of $\sim400 h^{-1}$ kpc. Hera also shows a bimodal structure, with a total mass of $\sim9.4\times10^{14} ~ M_{\odot}$ and a projected separation of $\sim130 h^{-1}$ kpc.   

We use 242 multiple images from 85 sources to reconstruct Ares, and 65 images from 19 sources for Hera. In addition, we include analytic nodes in the lens models. Degeneracies can arise between the grid and the analytic node components. 
To reduce degeneracy, the number of node components should be kept minimal.
Since our free-form reconstructions of both Ares and Hera reveal that each is composed of two cluster-scale mass structures \citep{2022ApJ...931..127C}, we employ two analytic nodes for each cluster.

\subsection{MACSJ0416 Data}\label{0416_data}

\begin{figure*}
\centering
\includegraphics[width=0.95\textwidth]{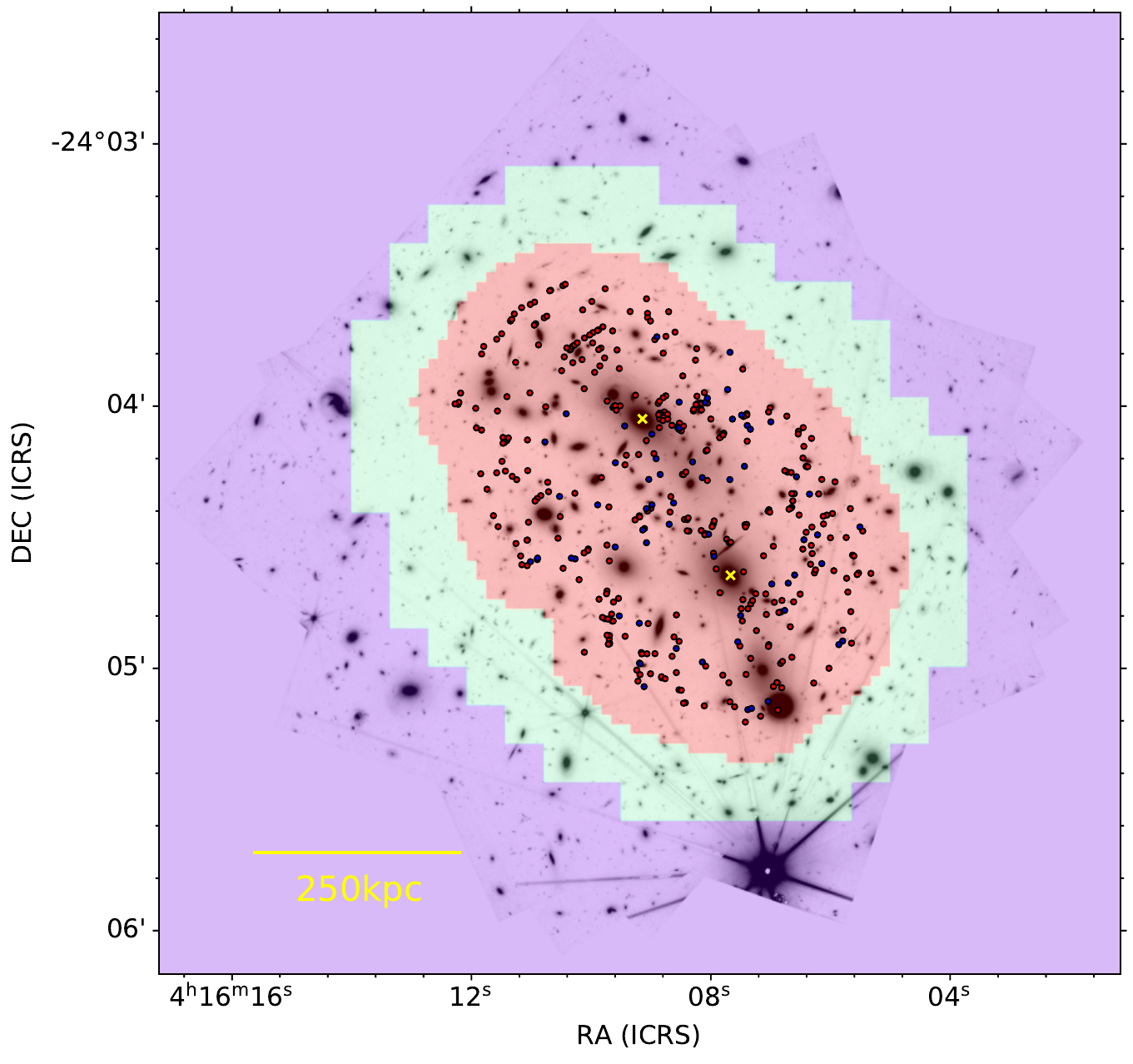} 
\caption{Multiple-image distributions and resolution levels for the reconstruction of MACSJ0416. The red (blue) circles indicate the positions of gold (silver) images, while the yellow crosses represent the initial locations of analytic nodes. The red, green, and purple shaded regions illustrate the resolutions of $1\farcs1$/pix, $2\farcs2$/pix, and $8\farcs8$/pix, respectively. The mosaic image is created using the F200W filter.}
\label{SL_input_data}
\end{figure*}

\subsubsection{JWST NIRCam Images}\label{JWST_images}
We use publicly available JWST NIRCam imaging data observed under the GTO programs 1176 (PI: Rogier A. Windhorst) and 1208 (PI: Chris J. Willott). 
We briefly outline our reduction process here. Data reduction followed the JWST pipeline \citep{Bushouse2024}, starting with the calibration reference file jwst\_1241.pmap\footnote{\url{https://jwst-crds.stsci.edu/}}. To correct snowball artifacts caused by cosmic ray hits, we applied the method described in \cite{Bagley2023}. Additionally, we applied the techniques outlined in \cite{Bagley2023} to remove 1/f noise and wisp artifacts. A custom flat-field image, generated from approximately 200 science images, was used to account for detector response variations. The processed images were then combined using the Drizzle algorithm with a pixel scale of  $0\farcs02$/pixel, yielding high-resolution outputs for further analysis.

\subsubsection{SL Data}\label{SL_data}
We adopt the multiple-image catalog published by the CANUCS team\footnote{\url{https://niriss.github.io/lensing.html}} \citep{2025A&A...696A..15R}, which compiles 415 multiple images in total, combining deep JWST/NIRCam and NIRISS observations with previous MUSE- and HST-based identifications. The catalog includes 38 newly identified images in 15 systems.
While \citet{2025A&A...696A..15R} constructed their lensing model using only the gold sample, which consists of 303 multiple images with spectroscopic redshifts, we use both the gold and silver samples, the latter of which do not have spectroscopic redshifts. 
Among the silver images, we exclude system K81, as its inclusion introduces unphysical fluctuations when the model forces its multiple images to converge to a level of scatter comparable to that of the others.
Figure~\ref{SL_input_data} presents the distribution of the 412 multiple images used in this study, consisting of 347 gold and 65 silver images.
The source redshifts of silver images are treated as free parameters and determined during the lens model optimization.

As with the simulated clusters, we prefer to keep the number of analytic nodes minimal to mitigate degeneracies between the grid and analytic components.
Motivated by our free-form reconstruction in \citet{2023ApJ...951..140C}, we place two analytic nodes on the northern and southern BCGs of MACS J0416, as indicated by the yellow crosses in Figure~\ref{SL_input_data}. 
We note that we tested a reconstruction with a larger number of analytic nodes (212 in total), as in \citet{2023A&A...674A..79B}, and found that although the overall mass distribution remained similar, two artificial (unphysical) peaks appeared near the northern BCG, indicating degeneracy between the grid and analytic components. Therefore, we adopted the two–analytic–node configuration for stability and interpretability.

\section{Method}\label{sec:method}
\subsection{Lensing Theory}
Here, we briefly summarize a lensing theory. For more details, we refer readers to review papers \citep[e.g.,][]{bartelmann2001, Kochanek2006, 2011A&ARv..19...47K, hoekstra2013, 2024SSRv..220...19N}.

The lens equation, which describes the relation between the source position $\bm{\beta}$ and the observed image position $\bm{\theta}$ is computed by the following: 
\begin{equation}\label{lens_equation}
    \bm{\beta}=\bm{\theta}-\bm{\alpha}(\bm{\theta}),
\end{equation}
where $\alpha$ indicates the deflection angle.
The deflection angle can be obtained through two different approaches. One of them is the derivative of the lensing potential $\varphi$:
\begin{equation}\label{eqn_deflection_via_poten}
    \bm{\alpha} = \nabla \varphi.
\end{equation}
Alternatively, the deflection angle can also be computed by convolving the convergence $\kappa$,
\begin{equation}\label{eqn_deflection_via_con}
    \bm{\alpha} (\bm{\theta}) = \frac{1}{\pi} \int
    \kappa (\bm{\theta}^{\prime}) \frac{\bm{\theta}-\bm{\theta}^{\prime}}{|\bm{\theta}-\bm{\theta}^{\prime}|^{2}} \bm{d {\theta}}^{\prime},
\end{equation}
where convergence $\kappa$ is the unitless surface mass density defined as:
\begin{equation}
    \kappa=\frac{\Sigma}{\Sigma_c}, ~ \Sigma_c = \frac{c^2 D_s}{4 \pi G D_d D_{ds}}.
\end{equation}
$\Sigma (\Sigma_c)$ is the (critical) surface mass density. $c$ and $G$ are the speed of light and gravitational constant, respectively. $D_{d(s)}$ indicates the angular diameter distance between the observer and lens (source), while $D_{ds}$ denotes the angular diameter distance between the source and the lens.

The lensing effect is a function of redshifts and the geometry of the Universe. Since the sources are at different redshifts, we have to consider their redshift differences. We first compute the deflection angle at a reference redshift $z_f$ and scale the deflection angle at the redshift of the source $z_s$ as follows:
\begin{equation}
    \bm\alpha(z_s) = \bm\alpha(z_f)W(z_f, z_s),
\end{equation}
where $W(z_f, z_s)$ is the angular diameter distances ratio:
\begin{equation}
    W(z_f, z_s) = \frac{D(z_f)D(z_l, z_s)}{D(z_s)D(z_f, z_l)}.
\end{equation}
In this paper, we set $z_f=\infty$ to simplify the calculation, and the deflection angle can be described by the following:
\begin{equation}\label{def_angle_redshift}
    \bm\alpha(z_s) = \frac{D(z_l, z_s)}{D(z_s)}\bm\alpha(z_f).
\end{equation}

\subsection{Truncated Pseudo-Elliptical NFW Profile}
{\tt MrMARTIAN} employs an analytic node in the lens modeling. In this study, we use truncated NFW profiles. Although the standard NFW profile \citep{1996ApJ...462..563N} has been widely used in both WL and SL analysis, the divergent total mass of the NFW profile is physically unrealistic and may introduce biases in the modeling process \citep{2011MNRAS.414.1851O}. 
Truncation of the analytic profile is usually used to describe tidal effects on galaxy-scale DM halos in dense environments such as galaxy clusters \citep[e.g.,][]{2007A&A...461..881L, 2009ApJ...696.1771L, 2017ApJ...841...18B}. 
Moreover, the truncated NFW profile ensures a finite total mass, addressing the divergent total mass of the standard NFW profile.
In this work, we follow the formulae described in \citet{2009JCAP...01..015B} to implement the smoothly truncated NFW profile.  

In addition to truncation, we also consider the ellipticity of the analytic nodes. However, since there is no general analytic expression for an elliptical NFW profile, we adopt a pseudo-elliptical approach. The pseudo-elliptical NFW profile has an elliptical lensing potential, not an elliptical convergence $\kappa$. In this paper, we adopt the description in \citet{2002A&A...390..821G} and \citet{2012A&A...544A..83D}. A brief description of the truncated NFW profile and the pseudo-elliptical approach is summarized in  Appendix~\ref{Apen_PETNFW}. For more details, we refer readers to the following papers: \citet[][]{2002A&A...390..821G, 2009JCAP...01..015B, 2012A&A...544A..83D}.

\subsection{{\tt MrMARTIAN}: Hybrid Lens Modeling Algorithm}\label{MARTIAN_method}

\begin{deluxetable}{cccccc}
\tablecaption{Prior ranges of the free parameters.
\label{tab:prior_range}
}
\tablehead {
\colhead{Parameter} &
\colhead{Prior} & 
\colhead{Unit} 
}
\startdata
$\kappa ~(\rm grid~  cell)$ & [-10, 10] & - \\
Halo position$^1$ ($x, ~ y$) & [-15, 15] & arcsecond \\
Concentration ($c$) & [0.01, 30] & -  \\
Scale radius ($r_s$) & [1, 1000] & kpc \\
Truncation ratio ($\tau$) & [0.01, 30] & - \\
Rotation angle ($i$) & [$-\infty$, $\infty$] & degree \\
Elliptical parameter ($\epsilon$) & [0, 0.25] & - \\
Model redshift ($z_{\rm model}$) & [0.496, 15] & -  
\enddata
\tablecomments{$^1$The coordinates are defined with respect to the position of each BCG.}
\end{deluxetable}

{\tt MrMARTIAN} reconstructs mass distributions by combining both grid cells and analytic nodes. The convergence $\kappa$ and deflection angle $\bm{\alpha}$ are computed as the sum of contributions from grid cells and analytic nodes:
\begin{align}
\begin{split}
    \kappa(\bm{\theta}) &= \kappa_{{\rm grid}}(\bm{\theta}) + \sum_{n=1}^N \kappa_{{\rm prof},n}(\bm{\theta}), \\
    \bm{\alpha} (\bm{\theta}) &= \bm{\alpha}_{\rm grid}(\bm{\theta}) + \sum_{n=1}^N \bm{\alpha}_{{\rm prof},n}(\bm{\theta}),
\end{split}
\end{align}
where the subscripts ``grid" and ``prof" indicate values of $\kappa$ and $\bm{\alpha}$ derived from grid cells and analytic profiles, respectively. $N$ is the total number of analytic profiles and $n$ represents the $n^{th}$ profile.
Each analytical node is described by seven free parameters: halo position $(x, y)$, concentration $c$, scale radius $r_s$, truncation ratio $\tau$, elliptical parameter $\epsilon$, and rotation angle $i$ measured counterclockwise from the x-axis. We set a flat prior for each parameter, as summarized in Table~\ref{tab:prior_range}. 

Unlike typical parametric methods, {\tt MrMARTIAN} optimizes halo parameters purely based on strong lensing data, without relying on optical priors such as scaling relations.
As noted in \textsection\ref{sec:data}, degeneracies can arise between the grid and analytic node components because their individual contributions to the total deflection cannot be uniquely separated.
Therefore, the risk of degeneracies would increase if we apply a large number of analytic halos.
To reduce the risk of degeneracy, we keep the number of analytic nodes minimal.

The {\tt MrMARTIAN} algorithm minimizes the following target function: 
\begin{equation}\label{total_eqn}
    f = \chi^2_{\rm SL} + rR, 
\end{equation}
where $\chi^2_{\rm SL}$ is the $\chi^2$ for SL and $R$ is the regularization term. The parameter $r$ determines the relative importance between $\chi^2_{\rm SL}$ and $R$.
$\chi^2_{\rm SL}$ is computed as 
\begin{equation}
    \chi^{2}_{\rm SL}=\sum_{i=1}^{I} \sum_{j=1}^{J}\frac{(\bm{\theta}_{i,j}-\bm{\alpha}_{i,j}(z_s)-\bm{\beta}_{i})^{2}}{{\sigma_{i}}^{2}},
\label{chi_square_SL}
\end{equation}
where
\begin{equation}
    \bm{\beta}_{i}=\frac{1}{J}\sum_{j=1}^{J}(\bm{\theta}_{i,j}-\bm{\alpha}_{i,j}(z_s)).
\end{equation}
$I$ and $J$ are the total number of SL image systems and the number of lensed images from each system, respectively. 
Here, $\sigma_i$ denotes a source-plane scatter derived from the dispersion among the delensed virtual-knot positions associated with image system $i$ (see Figure 1 in \citealt{2022ApJ...931..127C}). This approach discourages high-magnification (i.e., overfocusing) solutions in source-plane optimization.


For regularization, {\tt MrMARTIAN} improves upon the {\tt MARS} regularization method by allowing negative $\kappa$ values, following the scheme of \citet{1998MNRAS.298..905H}. 
This additional flexibility enables {\tt MrMARTIAN} to preserve the overall mass density profile set by the analytic components while accommodating local perturbations that are either lower or higher than those predicted by the profile. Specifically, the $\kappa$ value in each grid cell is decomposed into positive ($\kappa_p$) and negative ($-\kappa_n$) components (i.e., both $\kappa_p$ and $\kappa_n$ are positive), such that $\kappa_{\rm grid} = \kappa_p + (-\kappa_n)$, and the regularization term is computed as follows:
\begin{equation}
    R=\sum\left (p_{p} + p_{n} -\psi + \kappa_{\rm grid} \mathrm{ln} \frac{\psi + \kappa_{\rm grid}}{2 p_{p}}  \right),
\label{new_cross_entropy}
\end{equation}
where $p_{p} ~ (p_{n})$ indicates the prior of $\kappa_{p} ~ (\kappa_{n})$, and the summation is taken over pixels across all resolution scales. Similar to {\tt MARS}, {\tt MrMARTIAN} updates prior iteratively. At each minimization epoch, each prior value is evaluated at a single pixel after smoothing the $\kappa_{\rm grid}$ from the previous epoch with a Gaussian kernel of $\sigma=1.2$ grid cells.
$\psi$ is defined as:
\begin{equation}
    \psi = (\kappa_{\rm grid}^2 + 4 p_{p} p_{n})^{1/2}.
\end{equation}

For multiple images without spectroscopic redshifts, we consider their redshifts as free parameters in the modeling process. The deflection field is rescaled to each trial redshift by using Equation~\ref{def_angle_redshift}, and the target function $f$ in Equation~\ref{total_eqn} is minimized jointly over the mass-model parameters and these redshifts. As in \citet{2023ApJ...951..140C} and \citet{2024ApJ...961..186C}, we set a flat prior of $z_{\rm model}=[z_{\rm cluster}+0.1, 15]$, where $z_{\rm cluster}=0.396$ (see Table~\ref{tab:prior_range}). 
Unless stated otherwise, the presented $\kappa$ field is scaled to $D_{ds}/D_s=1$, whose corresponding critical surface mass density (i.e., $\kappa=1$) is $1.51\times10^9 M_{\odot} \rm{kpc}^{-2}$.

\subsection{Multi-resolution Approach}\label{multi_res_approach}
{\tt MrMARTIAN} employs a multi-resolution approach for lens modeling. By assigning a higher degree of freedom to regions with SL constraints, the number of free parameters can be reduced without significant loss of information. We apply a higher-resolution grid in regions strongly constrained by multiple images, and a lower-resolution grid where the constraints are weaker. At each iteration, the {\tt MrMARTIAN} algorithm oversamples low-resolution $\kappa$ regions to the highest resolution when calculating deflection angles via convolution. It then computes the model evaluation using Function~\ref{total_eqn} and updates parameters accordingly for the next iteration.

We apply three resolution levels to Ares and Hera. To avoid artifacts caused by large differences in pixel size, especially at the resolution boundaries, we include an intermediate level between the highest and lowest resolutions.
The same three-level resolution scheme is applied to the modeling of MACS J0416, using pixel scales of $1\farcs1$, $2\farcs2$, and $8\farcs8$.
Figure~\ref{SL_input_data} displays the regions corresponding to each resolution level. 
The outer (green and purple) regions beyond the multiple image positions serve as a margin in the deflection field calculation. This margin accounts for the potential influence of mass located outside the modeled FOV and helps suppress boundary artifacts that may arise during the convolution used to compute the deflection angle.
As a result, the number of free parameters for the $\kappa$ grid is reduced to 9928, which is $\mytilde13\%$ of the 78400 parameters required without the multi-resolution approach. 
The total number of free parameters used in the reconstruction is 9967: $N_{\rm grid}=9928$, $N_{\rm prof}=7\times2=14$, and $N_{\rm redshift}=25$.

\subsection{Uncertainty Estimation Method}\label{error_est}
\begin{deluxetable}{cccccc}
\tablecaption{Sampling ranges for the initial conditions.
\label{tab:initial_conditions}
}
\tablehead {
\colhead{Parameter} &
\colhead{Prior} & 
\colhead{Unit} 
}
\startdata
$\kappa~(\rm grid~  cell)$ & [-0.5, 0.5] & - \\
Halo position$^1$ ($x, ~ y$) & [-10, 10] & arcsecond \\
Concentration ($c$) & [0.01, 3] & -  \\
Scale radius ($r_s$) & [1, 100] & kpc \\
Truncation ratio ($\tau$) & [0.01, 30] & - \\
Rotation angle ($i$) & [0, 180] & degree \\
Elliptical parameter ($\epsilon$) & [0, 0.25] & - \\
Model redshift ($z_{model}$) & [0.496, 15] & -  
\enddata
\tablecomments{$^1$The coordinates are defined with respect to the position of each BCG.}
\end{deluxetable}

The {\tt MARS} algorithm estimates uncertainties by calculating the Hessian matrix of the target function $f$ in Equation~\ref{total_eqn} (see Section 2.2 in \citealp{2022ApJ...931..127C} for more details). While the Hessian-based approach is easy to implement and widely used \citep[e.g.,][]{1998MNRAS.299..895B, 2007ApJ...661..728J, 2022ApJ...931..127C}, it has some limitations. One limitation is the assumption that the posterior distributions of the free parameters follow a normal distribution. This assumption is generally invalid, especially for halo parameters \citep[e.g.,][]{2007NJPh....9..447J, 2009MNRAS.395.1319J, 2017ApJ...851...46F, 2019A&A...631A.130B, 2021ApJ...923..101K}. 

In this study, we estimate uncertainties by generating an ensemble of possible solutions, a technique commonly used in free-form methods \citep[e.g.,][]{2007MNRAS.375..958D, 2014ApJ...797...98L, 2019MNRAS.482.5666W, 2021MNRAS.506.6144G, 2025MNRAS.536.2690P}. Although ensemble-based approaches are more flexible and do not rely on Gaussian assumptions, they are computationally expensive. This made them an inefficient option for {\tt MARS} in its earlier versions. However, with the current improvement—most notably the significant reduction in the number of free parameters—such an approach has now become feasible within the {\tt MrMARTIAN} framework.
We reconstruct 200 realizations from randomly generated initial conditions, sampled within the ranges listed in Table~\ref{tab:initial_conditions}.

\section{Result} \label{sec:result}
\subsection{Mock Cluster Result}
To investigate the effect of employing analytic nodes for the reconstruction, we compare the lens models of Ares and Hera using {\tt MrMARTIAN} with the ground truth maps, as well as with results by {\tt MARS}. 
In this paper, we adopt a $\kappa=0$ grid as the initial condition \footnote{We note that \citet{2022ApJ...931..127C} uses different initial conditions used in this work: a field with $\kappa=0.5$.}.
We display the central regions of $180\arcsec\times180\arcsec$ for Ares and $110\arcsec\times110\arcsec$ for Hera. The maps are resampled at $0\farcs14$/pixel and $0\farcs11$/pixel for Ares and Hera, corresponding to the resolution of the publicly available data.

\subsubsection{Lens Model Comparison}
\label{mock_lens_comparison}
\begin{figure*}
\centering
\includegraphics[width=\textwidth]{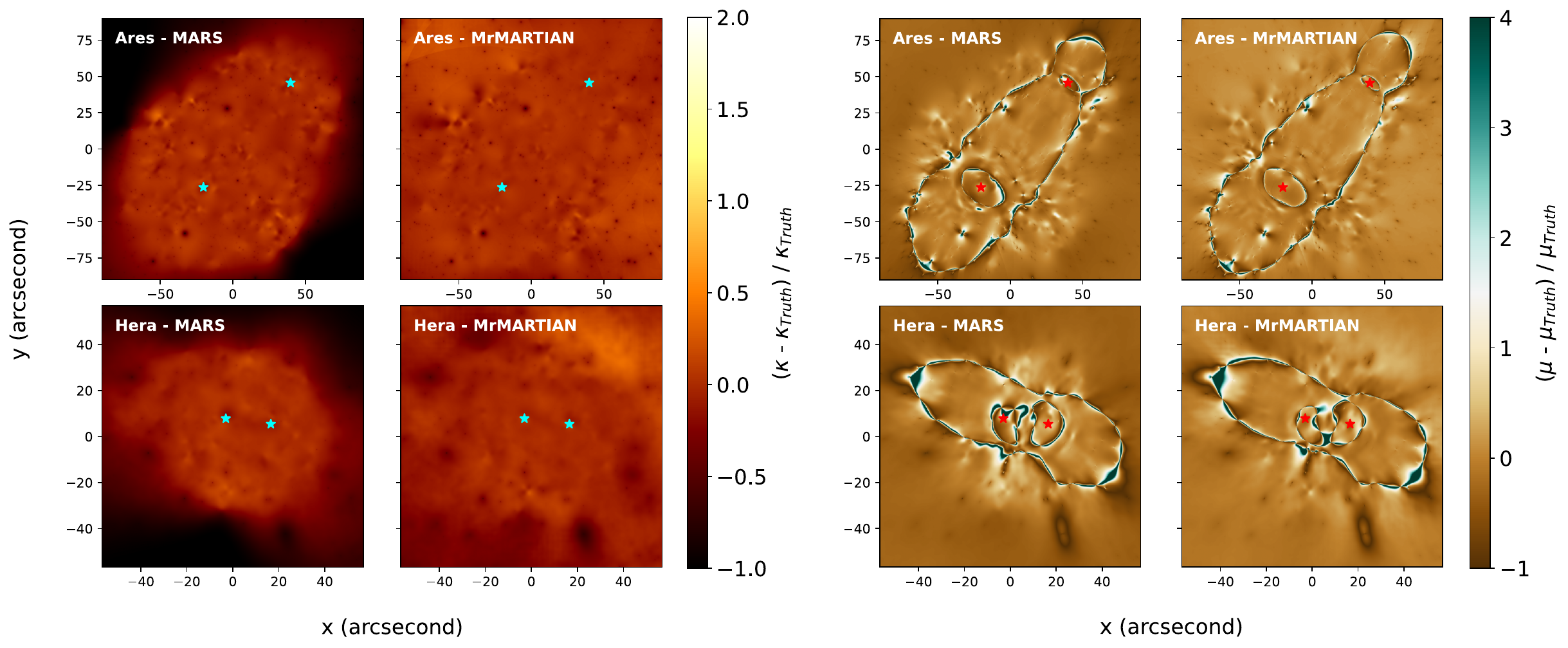} 
\caption{Relative difference between the true and reconstructed maps. The panels on the left (right) side indicate the residual maps of the convergence (magnification). The cyan (red) markers represent the locations of BCGs. The median relative differences of $\kappa$ are -0.125 and -0.350 (0.017 and 0.054), for Ares and Hera from {\tt MARS} ({\tt MrMARTIAN}), respectively.}
\label{mock_kappa_map_difference}
\end{figure*}

\begin{figure*}
\centering
\includegraphics[width=\textwidth]{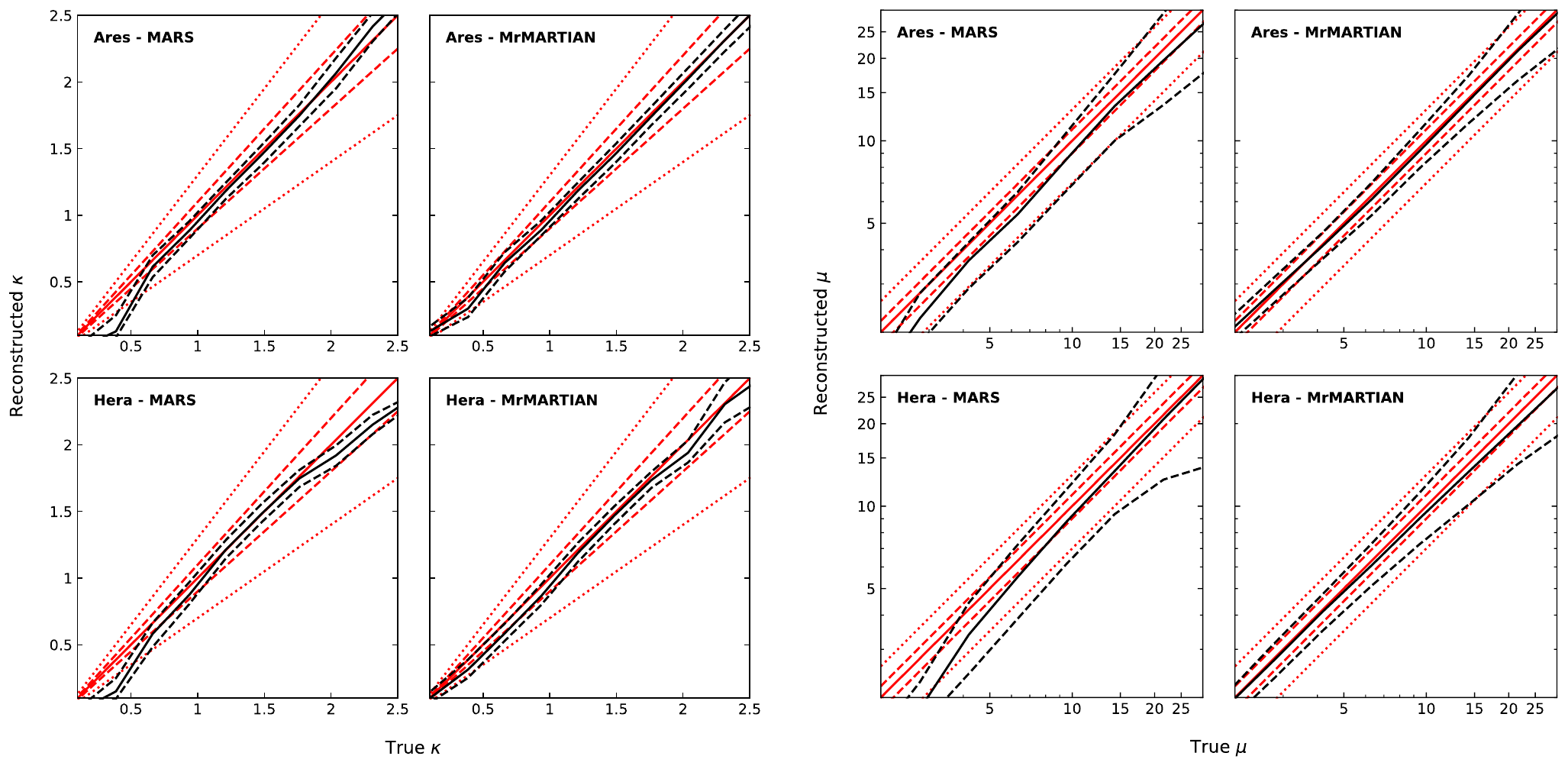} 
\caption{Convergence (left) and magnification (right) correlation between the true and reconstructed lens models. The black solid (dashed) lines correspond to the median (25th and 75th percentiles). The red solid lines show the perfect correlation. The red dashed (dotted) lines present indicate $\pm10 \%$ ($\pm30 \%$) deviations from the true values.}
\label{one2one_comp}
\end{figure*}

\begin{figure}
    \includegraphics[width=0.4\textwidth]{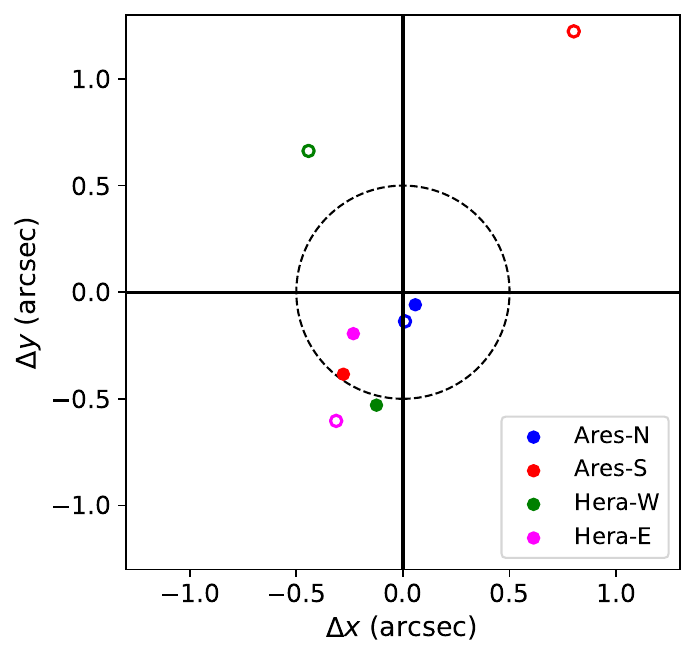} 
    \caption{Mass peak offsets between the true and reconstructed mass models. The filled (empty) circles indicate the offsets derived from {\tt MrMARTIAN} ({\tt MARS}). The black dashed line shows the $0\farcs5$ offsets from the true mass peaks.} 
\label{peak_offset}
\end{figure}

In Figure~\ref{mock_kappa_map_difference}, we compare the relative differences between the true and reconstructed mass and magnification maps for both synthetic clusters using the full SL catalog. 
While the reconstructed mass maps from {\tt MARS} and {\tt MrMARTIAN} show small differences within the SL regime, significant discrepancies are observed in the cluster outskirts. 
Since there are no SL multiple images, the convergence $\kappa$ from {\tt MARS} mainly relies on its initial prior. 
By comparison, {\tt MrMARTIAN} recovers the convergence well in both the outskirts and central regions, thanks to the inclusion of analytic halo profiles.
Similar trends appear in the magnification maps, where {\tt MrMARTIAN} more accurately reproduces the true values than {\tt MARS} in outer regions, despite smaller discrepancies than in the convergence maps.

In Figure~\ref{one2one_comp}, we present quantitative assessments by comparing the convergence and magnification values in the same regions.
For convergence, the {\tt MARS} result is clearly lower than the true values in the low-$\kappa$ ($\lesssim 0.6$) regime, mainly due to the underestimation observed in the outskirts of the field (Figure~\ref{mock_kappa_map_difference}). 
In the high-$\kappa$ ($\gtrsim 2$) regime, the {\tt MARS} result underestimates $\kappa$ for Hera, likely caused by regularization-induced oversmoothing of the mass peaks.
This bias is absent in the Ares result, possibly owing to denser SL constraints near the mass peaks (see Figures 2 and 3 of \cite{2022ApJ...931..127C}).
Notably, {\tt MrMARTIAN} yields more accurate results across the entire $\kappa$ range, with median values deviating from the truth by $\lesssim 10\%$.
The difference in magnification between the {\tt MARS} and {\tt MrMARTIAN} results is less pronounced than in convergence; nevertheless, the {\tt MrMARTIAN} results remain superior in both precision and accuracy.

Another metric for assessing the quality of mass reconstruction is the accuracy of mass peak positions. We find that {\tt MrMARTIAN} shows improved performance in this aspect.
In Figure~\ref{peak_offset}, we compare the offsets between the reconstructed and true mass peak positions for Ares and Hera.
Of the four mass peaks, {\tt MrMARTIAN} places three within $0\farcs5$ of the true positions and the remaining one within $0\farcs6$, whereas {\tt MARS} places only one within $0\farcs5$, with the largest offset reaching $\gtrsim1\arcsec$.
Although the comparison is based on a small sample of two, we expect the trend to hold more generally, as {\tt MrMARTIAN} benefits from the assumption that highly concentrated mass peaks are located near the BCGs.

\subsubsection{Effect of Multiple-Image Sparsity}
\begin{figure*}
\centering
\includegraphics[width=\textwidth]{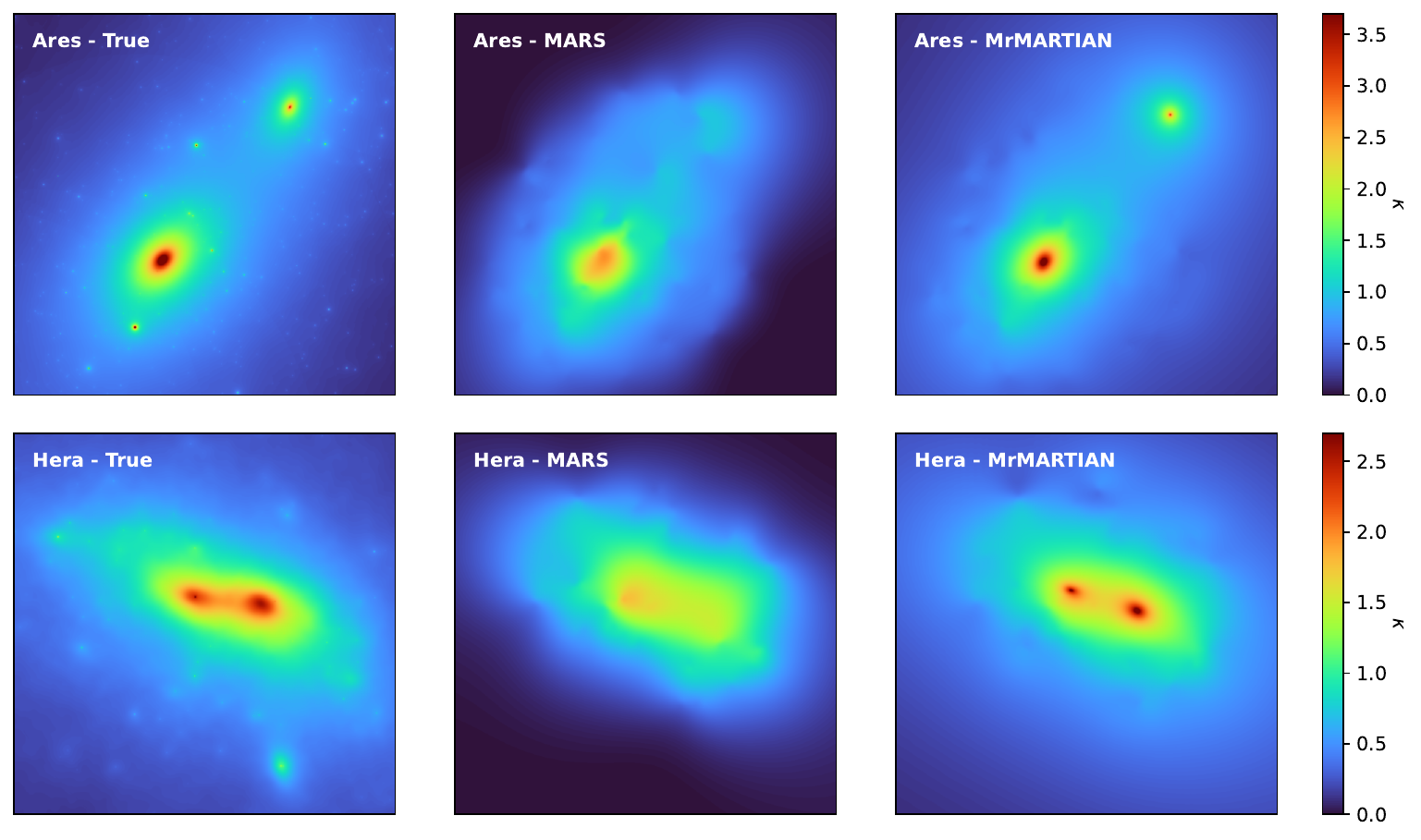} 
\caption{Reconstructed mass of Ares and Hera with sparsely distributed SL multiple images. We randomly select 54 (35) SL multiple images from 242 (65) multiple images of Ares (Hera). Left: true mass maps. Middle: mass maps reconstructed using {\tt MARS}. Right: reconstructed mass maps from {\tt MrMARTIAN}.} 
\label{multi_image_compare}
\end{figure*}

One advantage of integrating analytic nodes is the ability to achieve stable reconstructions even with a small number of multiple images.
In general, pure free-form methods are more susceptible to the sparsity of multiple images than parametric ones, as the system is more underdetermined relative to the number of free parameters.
We anticipate that the hybrid approach implemented in {\tt MrMARTIAN} effectively mitigates this limitation.

To test this, we randomly select 54 (35) SL multiple images from the full catalog of 242 (65) images from Ares (Hera). Figure~\ref{multi_image_compare} shows the resulting mass reconstructions.
While sparsity degrades the reconstruction quality in both Ares and Hera, the {\tt MARS} results are more severely affected, exhibiting overly smoothed mass peaks. In some cases, the weak mass peaks are difficult to identify altogether.

\subsection{MACSJ0416 Result}\label{0416_results}
{\tt MrMARTIAN} yields quasi-unique solutions across different initial conditions; however, some differences in fine details are inevitable. In this section, we present the result obtained using the following initial conditions, which we refer to as the reference model hereafter:
$(\kappa, c, r_s, \tau, i, \epsilon, z_{model}) = (0, 0.01, 1, 0.01, 0, 0, 3)$, 
with the two analytic halos at the positions of BCG-N and BCG-S. 
As with both Ares and Hera, the multi-resolution approach is also employed for the reconstruction of MACSJ0416. 
For the final presentation, the entire mass reconstruction result is resampled using bicubic interpolation at $0\farcs02$ per pixel, matching the pixel scale of our JWST mosaic image.

\subsubsection{Mass Model}\label{mass_maps}
\begin{figure*}
\centering
\includegraphics[width=0.95\textwidth]{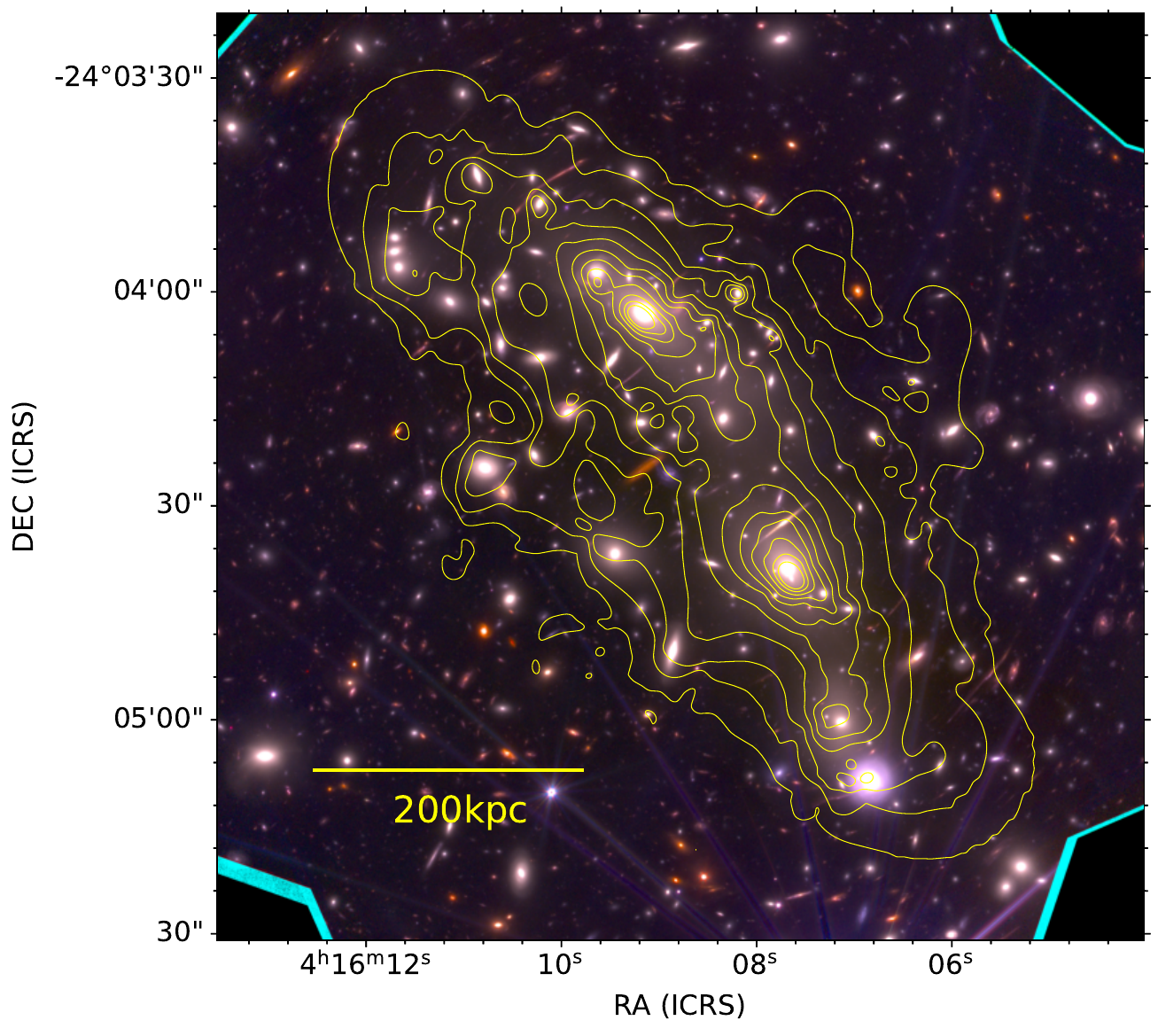} 
\caption{Mass contours of MACSJ0416 overlaid on the color-composite image. The yellow contours show the convergence $\kappa$ and range from 0.6 to 2.1 with intervals of 0.15. The color-composite image is generated using the F090W + F115W filters for blue, the F150W + F200W + F277W filters for green, and the F356W + F444W filters for red.}
\label{result_kappa_color_image}
\end{figure*}

\begin{figure*}
\centering
\includegraphics[width=0.95\textwidth]{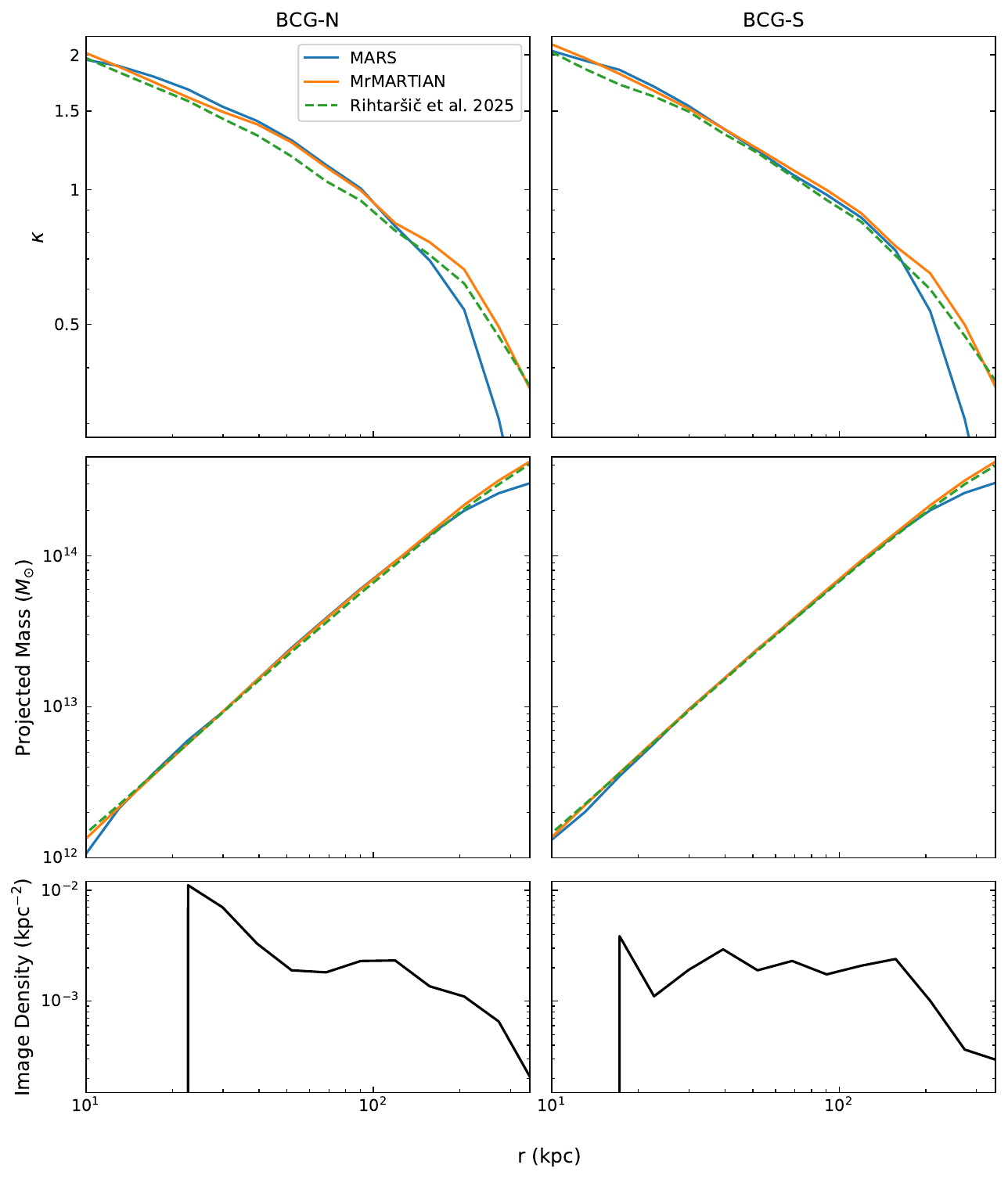} 
\caption{Radial $\kappa$ (top), projected mass (middle), and SL multiple image density (bottom) profile comparison. We display radial profiles of BCG-N and BCG-S. The solid blue and dashed orange lines indicate the profiles derived from the {\tt MARS} and {\tt MrMARTIAN}, respectively. The solid green line represents the profiles derived from the \citet{2025A&A...696A..15R}. The solid black lines in the bottom panels show the SL multiple image density.}
\label{radial_profile}
\end{figure*}

Figure~\ref{result_kappa_color_image} shows the reconstructed mass map of MACSJ0416. Our mass model reveals the elongated mass along the northeast-southwest direction, consistent with the result from {\tt MARS} \citep[see Figure 4 in][]{2023ApJ...951..140C} and other previous studies \citep[e.g.,][]{2023A&A...674A..79B, 2024A&A...681A.124D, 2025A&A...696A..15R, 2025MNRAS.536.2690P}.
To assess the performance of our mass map, we evaluate the root-mean-square (rms) of the position differences between the observed and predicted multiple images: 
\begin{equation}\label{eqn_rms}
    \Delta_{\rm rms}=\sqrt{\frac{1}{M}\sum_{m=1}^{M}|\bm{\theta}_{truth,m}-\bm{\theta}_{model,m}|^{2}},
\end{equation}
where $M$ indicates the number of multiple images, while $\bm{\theta}_{model,m}$ and $\bm{\theta}_{truth,m}$ represent the locations of the predicted and observed multiple images for the $m^{th}$ image, respectively.

The rms value of our mass model is $0\farcs11$, $\mytilde25\%$ larger than the rms value from {\tt MARS} \citep[][$\Delta_{\rm rms}=0\farcs084$]{2023ApJ...951..140C}. This increase is not surprising, as we use approximately twice as many multiple images \citep[][236 images]{2023ApJ...951..140C}, and some of the newly added images do not have secure redshift information. However, we note that our rms is still smaller than the values from other results using both parametric and free-form methods. 
Using a parametric method, \citet[]{2023A&A...674A..79B} achieved an rms value of $\Delta_{\rm rms} = 0\farcs43$ based on 237 multiple images from HST observations. A slightly higher rms of $0\farcs53$ was reported by \citet[]{2025A&A...696A..15R}, who used 303 multiple images from JWST observations.
\citet{2025MNRAS.536.2690P}, using the free-form algorithm {\tt GRALE} \citep{2006MNRAS.367.1209L, 2007MNRAS.380.1729L}, report an rms scatter of $0\farcs19$ based on 237 spectroscopically confirmed multiple images, without incorporating JWST observations.
The small rms value ($0\farcs11$) in our mass model demonstrates the ability of {\tt MrMARTIAN} to reconstruct mass density profiles using analytic nodes while preserving the flexibility of {\tt MARS}.

Figure~\ref{radial_profile} compares the cumulative mass and radial $\kappa$ profiles for BCG-N and BCG-S among three different mass models: {\tt MARS}, {\tt MrMARTIAN}, and the model from \citet{2025A&A...696A..15R}, who make their results publicly available. \citet{2025A&A...696A..15R} reconstructs the mass model of MACSJ0416 using {\tt lenstool} \citep{2007NJPh....9..447J}, based on 303 spectroscopically confirmed multiple images identified in JWST observations. 

The radial profiles from {\tt MARS} are consistent with {\tt MrMARTIAN}  within the SL regime ($\lesssim150$ kpc). 
Beyond this radius, as the density of multiple images decreases, the convergence $\kappa$ drops rapidly. 
As discussed in \textsection\ref{mock_lens_comparison}, this decline results from the lack of SL constraints, causing {\tt MARS} to default to the flat initial prior.
When compared with the lens model from \citet{2025A&A...696A..15R}, {\tt MrMARTIAN} yields slightly higher $\kappa$ values in both the BCG-N and BCG-S regions.
Given the differences in the selection of multiple images and the mass modeling algorithm, this mass offset is small, corresponding to $\mytilde3$\% discrepancy in the projected mass at $r=200$ kpc.

Recently, \citet{2025arXiv250616034L} investigated the small mass feature ``M2" that had been reported in some previous reconstructions of MACSJ0416. Using both models from {\tt lenstool} \citep{2025A&A...696A..15R} and {\tt GRALE} \citep{2025MNRAS.536.2690P}, \citet{2025arXiv250616034L} found that the parametric model does not require any explicit clump at the M2 location, while the free-form model shows only a weak residual signal. We also examined our convergence map and found no indication of a localized mass peak near the M2 position, consistent with their findings. This result implies that M2 is unlikely to represent a genuine DM substructure, and further supports that DM substructures are generally associated with luminous counterparts, consistent with the light-affiliated nature of DM predicted by the $\Lambda$CDM paradigm.

\subsubsection{Magnification}\label{magnification}
\begin{figure*}
\centering
\includegraphics[width=\textwidth]{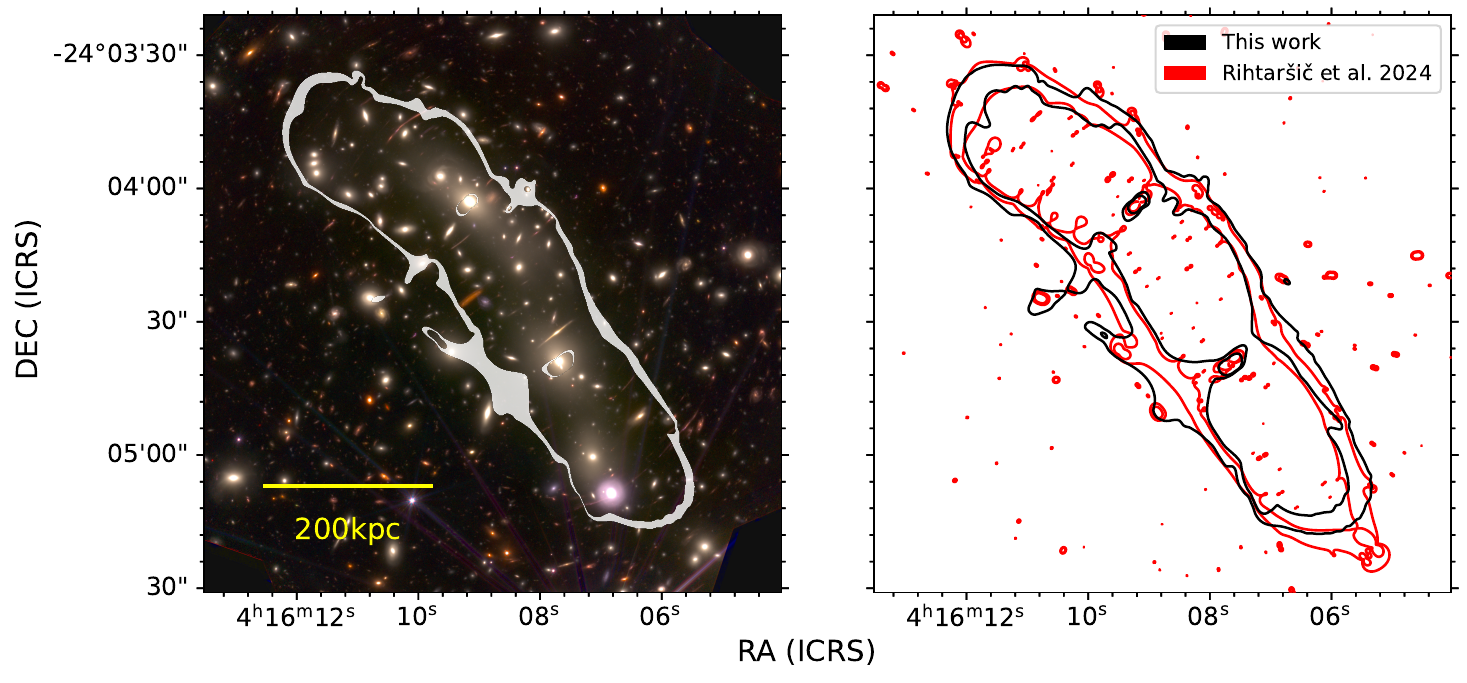} 
\caption{Magnification maps of MACSJ0416 at the reference redshift $z_f=9$. The white curves in the left panel indicate regions where the magnification is larger than 50. The black (red) contours in the right panel represent the magnification from this study (from \citealp{2025A&A...696A..15R}). The magnification contours indicate $|\mu|=20$ at the reference redshift $z_f=9$. The color-composite image is the same as in Figure~\ref{result_kappa_color_image}.}
\label{magnification_map}
\end{figure*}

In Figure~\ref{magnification_map}, we present the magnification map derived from our lens model. It features three major loops: two inner loops surrounding each BCG and one outer loop. While the overall morphology agrees with results in the literature, some details differ. When compared with the result from \citet{2025A&A...696A..15R} (see right panel), our magnification map lacks many small loops around the cluster galaxies because we assign analytic halos only to the two BCGs, whereas  \citet{2025A&A...696A..15R} assign them to every cluster member.
Furthermore, their critical curve extends farther to the southwest than in our model, owing to the inclusion of galaxies in that region in their modeling.

\section{Discussion} \label{sec:discuss}
\subsection{Effect of Multi-resolution}\label{multi_res_effect}
\begin{figure*}
\centering
\includegraphics[width=\textwidth]{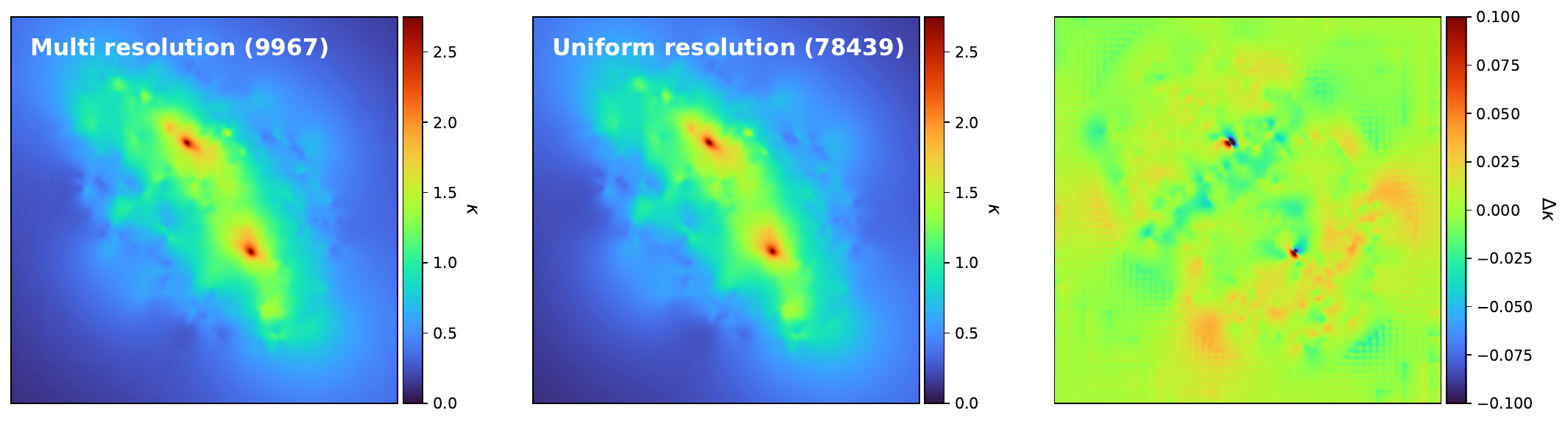} 
\caption{Comparison between the uniform and multi-resolution. The left (middle) column shows the mass reconstruction with (without) the multi-resolution approach. The number on the upper left in each panel presents the number of free parameters for lens modeling.
The right column represents the residual map between the mass maps with and without a multi-resolution approach.} 
\label{multi_res_compare}
\end{figure*}

As described in \textsection\ref{multi_res_approach}, {\tt MrMARTIAN} adopts a multi-resolution approach for reconstruction. While this enhances the modeling efficiency, the use of multiple resolutions can introduce biases in the reconstructed mass maps due to resolution mixing.
To assess the potential bias of the multi-resolution approach, we compare reconstructions obtained with uniform and multi-resolution grids. The resolution of the uniform grid model is set to match the highest resolution used in the multi-resolution model. Any significant deviation of the multi-resolution result from the uniform grid result would indicate the presence of biases.

Figure~\ref{multi_res_compare} compares the mass maps with the uniform grid and multi-resolution models. The two results are in excellent agreement across the entire field, with no significant discrepancies apparent upon visual inspection. The residual map (right panel) shows low-contrast ($\Delta \kappa\sim0.01$) grid patterns in the cluster outskirts, where the multi-resolution model employs low-resolution grids. This is inevitable, as differences in grid cell size produce distinct pixelation effects.
In addition, some large-scale residual patterns perpendicular to the elongation of the MACS0416 system are visible. The maximum amplitude of these residuals is $\mytilde0.025$ which is comparable to the mass reconstruction uncertainties (see Figure~\ref{error_result}). However, since the average difference lies well below the uncertainty level, we conclude that both reconstructions are highly consistent. Their radial $\kappa$ profiles are also nearly indistinguishable. 
Finally, we note that the large differences at the BCG locations should not be interpreted as evidence of bias, as the positions of the analytic halos can vary between different mass reconstruction runs.

The above test shows that {\tt MrMARTIAN} maintains reconstruction quality while using only $\mytilde13\%$ of free parameters required by a high-resolution uniform grid. This capability enables faster and more efficient computation of both the lens model and its associated uncertainties than our previous uniform-resolution approach.

\subsection{Mass Reconstruction Uncertainty}\label{error_effect}
\begin{figure}
    \includegraphics[width=0.47\textwidth]{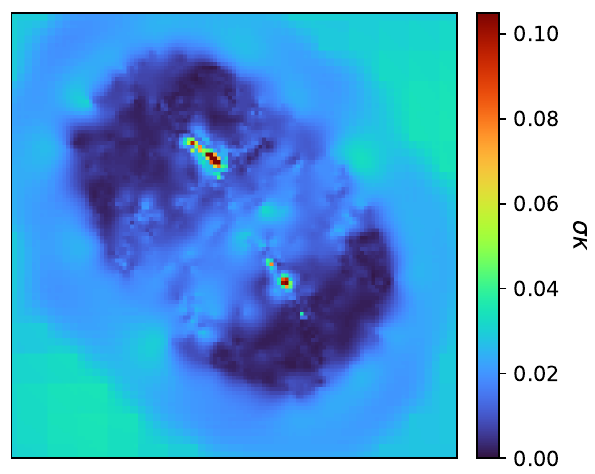} 
    \caption{The standard deviation of the 200 realizations. The resolution of the $\kappa$ pixels is $1\farcs1/$pixel.} 
\label{error_result}
\end{figure}

\begin{deluxetable*}{cccccccc}\label{analytic_node_results}
\tablecaption{Uncertainties of the two analytic nodes.}
\tablehead {
\colhead{Analytic Node} &
\colhead{$x^1~(\arcsec)$} &
\colhead{$y^1~(\arcsec)$} &
\colhead{$c$} & 
\colhead{$r_s~({\rm kpc})$} &
\colhead{$\tau^3$} &
\colhead{$i ~({\rm degree})$} &
\colhead{$\epsilon$}
}
\startdata
BCG-N$^2$ & $-0.13^{+0.27}_{-0.63}$ & $-0.34^{+0.54}_{-0.34}$ & $4.11^{+0.38}_{-0.42}$ & $350^{+56}_{-40}$ & $19.87^{+8.83}_{-11.64}$ & $138.64^{+6.26}_{-5.18}$ & $0.248^{+0.001}_{-0.042}$ \\
BCG-S$^2$ & $0.48^{+0.13}_{-0.12}$ & $-0.37^{+0.14}_{-0.16}$ & $4.19^{+0.32}_{-0.41}$ & $359^{+47}_{-40}$ & $17.90^{+10.79}_{-9.45}$ & $130.13^{+4.63}_{-2.80}$ & $0.249^{+0.001}_{-0.002}$ 
\enddata
\tablecomments{$^1$Offsets relative to each BCG. $^2$ Median (center), 16th (lower), and 84th (upper) percentiles from the 200 realization. $^3$ Truncation ratio is used only to control the smooth outer cut-off of the TNFW profile and is not interpreted as a physical tidal radius.
We note that the median values are not derived from the reference model.
}
\end{deluxetable*}

In Figure~\ref{error_result}, we present the standard deviation from the 200 realizations using randomly generated initial conditions. Across the entire FOV, the error of the mass map is $\sigma\kappa < 0.05$, except around the BCGs. The relatively larger uncertainties around the BCGs ($\sigma\kappa\sim0.1$) are mainly caused by the dispersion in the positions of the analytic nodes. 
Table~\ref{analytic_node_results} presents the uncertainties in the parameters of the two analytic nodes derived from the 200 realizations. We note that the median values in Table~\ref{analytic_node_results} are not those of the reference model. Although the halo positions are allowed to move freely, their uncertainties remain below $\sim0.5\arcsec$. All parameters except for the truncation ratio are consistent within $1\sigma$ between the two analytic nodes. The truncation ratios show large uncertainties, which span most of the prior range. This is not surprising, as the truncation ratio affects the outskirts of the analytic profiles where there are no SL constraints. We note that $\tau$ is used only to control the smooth outer cut-off of the TNFW profile and is not interpreted as a physical tidal radius.
Despite this, the overall mass distributions across the 200 realizations remain consistent, supporting that the large uncertainties in the truncation ratios have a limited impact on the mass reconstruction within the SL regime.
In addition, the variation of the deflection angle with respect to $\tau$ is not significant in the SL regime. \citet{2009JCAP...01..015B} showed that the difference between models with $\tau=10 {\rm ~ and} ~ 20$ remains small within $\sim2r_s$ (see their Figure 3).

\section{Conclusion} \label{sec:conclusion}
By combining grid cells and analytic nodes, we have presented a new hybrid SL modeling algorithm {\tt MrMARTIAN}. The grid component is regularized through maximum cross-entropy, as in {\tt MARS}, preserving flexibility and smoothness in mass map reconstruction. {\tt MrMARTIAN} also supports negative $\kappa$ values for lens modeling. To address the limitations of {\tt MARS}, which tends to oversmooth mass peaks and has difficulty reproducing the global mass density profile beyond the SL regime, we introduce the pseudo-elliptical truncated NFW profile as an analytic node in this study. 

We validate {\tt MrMARTIAN} using the publicly available simulated galaxy clusters Ares and Hera. {\tt MrMARTIAN} reconstructs global mass density profiles better than {\tt MARS} beyond SL regimes, where no multiple images exist. For magnification, {\tt MrMARTIAN} shows good agreement with true values while {\tt MARS} exhibits large deviations. In scenarios with sparsely distributed multiple images, {\tt MrMARTIAN} shows better performance than {\tt MARS}, effectively recovering compact core structures.

In addition, we apply {\tt MrMARTIAN} to MACSJ0416, using 412 multiple images. Our resulting mass map reveals two prominent density peaks at the two BCGs with flexible mass distributions indicated by the small scatters in the image plane ($\mytilde0\farcs11$). The critical curve from our lens model differs in the southwest regions from the previous parametric lens models. We attribute this discrepancy to the absence of multiple images and different analytic halo implementations between lens modeling algorithms. 

We assess the robustness of the multi-resolution implementation in {\tt MrMARTIAN}.
The deviations are observed between the uniform and multi-resolution near the boundaries of different resolution levels, with differences comparable to the uncertainties in convergence $\kappa$.
However, despite these discrepancies, the radial convergence $\kappa$ and cumulative projected mass remain consistent with the uniform-resolution result. In the MACSJ0416 case, the multi-resolution approach reduces free parameters to $\sim13\%$ of those required for uniform resolution, enabling faster and more efficient computation without losing information.

In this study, we demonstrate that {\tt MrMARTIAN} can reconstruct flexible mass distributions with robust density profiles, using unprecedentedly large SL multiple images. This capability enables us to reveal detailed mass distributions, the nature of DM, and the evolution of high-redshift galaxies in the JWST era.

We express our gratitude to Hyungjin Joo for providing the JWST NIRCam image of MACSJ0416. This work is based on observations created with NASA/ESA/CSA JWST and downloaded from the Mikulski Archive for Space Telescope (MAST) at the Space Telescope Science Institute (STScI).
The specific observations analyzed can be accessed via \dataset[doi: 10.17909/v0rx-d363]{http://dx.doi.org/10.17909/v0rx-d363}. M. J. Jee acknowledges support for the current research from the National Research Foundation (NRF) of Korea under the programs 2022R1A2C1003130 and RS-2023-00219959. SC acknowledges this research was supported by Basic Science Research Program through the NRF funded by the Ministry of Education (No. RS-2024-00413036).

\software{Astropy \citep{astropy2013,2018AJ....156..123A,2022ApJ...935..167A}, Matplotlib \citep{matplotlib2007}, NumPy \citep{harris2020array}, PyTorch \citep{NEURIPS2019_bdbca288}, SciPy \citep{scipy2020}}

\appendix
\section{Truncated Pseudo-Elliptical NFW Profile}\label{Apen_PETNFW}
\subsection{Truncated NFW Profile}
In this section, we briefly explain the truncated NFW profile. For more details, we refer readers to \citet{2009JCAP...01..015B}. The NFW profile \citep{1996ApJ...462..563N} is one of the most widely used universal profiles to describe DM distributions and is formulated by the following: 
\begin{equation}
    \rho(r)=\frac{\delta_c \rho_c}{(r/r_s)(1+r/r_s)^2},
\end{equation}
where $\rho_c$ and $r_s$ are the critical density and the scale radius, respectively. The characteristic overdensity $\delta_c$ is defined as:
\begin{equation}
    \delta_c=\frac{\Delta}{3}\frac{c^3}{{\rm ln}(1+c) - c/(1+c)},
\end{equation}
where $c$ is the concentration and $\Delta$ indicates the density contrast. In this study, we use $\Delta=200$.

The truncated NFW profile is designed to implement a smooth truncation after the tidal radius of the DM halo. The smoothed truncation of the NFW profile is computed as follows:
\begin{equation}
    \rho_T(x)=\frac{\delta_c \rho_c}{x(1+x)^2}(\frac{\tau^2}{\tau^2+x^2})^n,
\end{equation}
where $n$ controls the degree of truncation while $x$ indicates $r/r_s$. $\tau$ represents $r_t/r_s$, where $r_t$ is the tidal radius. In this study, we set $n=1$ to implement the truncation of the NFW profile.

\subsection{Pseudo-Elliptical Profile}
We refer readers to \citet{2002A&A...390..821G, 2012A&A...544A..83D} for more details.
In this paper, we adopt the following description to implement the pseudo-elliptical profile: 
\begin{align}
\begin{split}
    a_{1\epsilon} = 1-\epsilon, \\
    a_{2\epsilon} = 1+\epsilon,
\end{split}
\end{align}
where the elliptical parameter $\epsilon$ is in range $0\leq\epsilon<1$ \citep{1987ApJ...321..658B}. We set the upper limit of the elliptical parameter to 0.25 to avoid the unphysical peanut or boxy shapes of the mass distribution in this study \citep{2002A&A...390..821G, 2012A&A...544A..83D}.
The ellipticity of the lensing potential, $\epsilon_\varphi$ is defined as:
\begin{equation}
    \epsilon_\varphi = 1 - \sqrt{\frac{a_{1\epsilon}}{a_{2\epsilon}}} = 1 - \sqrt{\frac{1-\epsilon}{1+\epsilon}}.
\end{equation}
For dimensionless coordinates, we define $\mathbf{x}=\mathbf{r}/r_s$ where $\mathbf{r}$ is the distance from the center of a profile in the lens plane. An elliptical coordinate system can be expressed by the following:
\begin{equation}
\begin{cases}
    x_{1\epsilon} &= \sqrt{a_{1\epsilon}} x_1 \\
    x_{2\epsilon} &= \sqrt{a_{2\epsilon}} x_2 \\
    x_{\epsilon} &= \sqrt{x_{1\epsilon}^2 + x_{2\epsilon}^2} \\
    \phi_{\epsilon} &= {\rm arctan} (x_{2\epsilon}/x_{1\epsilon})
\end{cases}
\end{equation}
where $x_{1(2)}$ is the first (second) component of $\mathbf{x}$. 
From the above expressions, $\kappa_\epsilon$, $\alpha_\epsilon$, $\gamma_\epsilon$, and $\varphi_\epsilon$ derived from a pseudo-elliptical NFW profile are
\begin{equation}
\begin{cases}
    \kappa_{\epsilon}(\mathbf{x}) &= \kappa(x_{\epsilon}) + \epsilon ~ {\rm cos} 2\phi_{\epsilon} \gamma(x_{\epsilon}) \\
    \alpha_{1 \epsilon}(\mathbf{x}) &= \alpha(x_{\epsilon}) \sqrt{a_{1\epsilon}} {\rm cos} \phi_{\epsilon} \\
    \alpha_{2 \epsilon}(\mathbf{x}) &= \alpha(x_{\epsilon}) \sqrt{a_{2\epsilon}} {\rm sin} \phi_{\epsilon} \\
    \gamma_{1 \epsilon}(\mathbf{x}) &= - \epsilon ~ \kappa(x_{\epsilon}) - \gamma(x_{\epsilon}) {\rm cos} 2\phi_{\epsilon} \\
    \gamma_{2 \epsilon}(\mathbf{x}) &= - \sqrt{1 - \epsilon^2} ~ \gamma(x_{\epsilon}) {\rm sin} 2\phi_{\epsilon} \\
    \varphi_{\epsilon}(\mathbf{x}) &= \varphi(x_{\epsilon})
\end{cases}
\end{equation}
where $\kappa, ~ \alpha, ~ \gamma$, and $\varphi$ are the expressions of convergence, deflection angle, shear, and lensing potential in the simple NFW halo profile in the dimensionless coordinates \citep{1996A&A...313..697B}.

\bibliographystyle{aasjournal}
\bibliography{main}

@ARTICLE{astropy2013,
       author = {{Astropy Collaboration} and {Robitaille}, Thomas P. and
         {Tollerud}, Erik J. and {Greenfield}, Perry and {Droettboom}, Michael and
         {Bray}, Erik and {Aldcroft}, Tom and {Davis}, Matt and
         {Ginsburg}, Adam and {Price-Whelan}, Adrian M. and
         {Kerzendorf}, Wolfgang E. and {Conley}, Alexander and {Crighton}, Neil and
         {Barbary}, Kyle and {Muna}, Demitri and {Ferguson}, Henry and
         {Grollier}, Fr{\'e}d{\'e}ric and {Parikh}, Madhura M. and
         {Nair}, Prasanth H. and {Unther}, Hans M. and {Deil}, Christoph and
         {Woillez}, Julien and {Conseil}, Simon and {Kramer}, Roban and
         {Turner}, James E.~H. and {Singer}, Leo and {Fox}, Ryan and
         {Weaver}, Benjamin A. and {Zabalza}, Victor and {Edwards}, Zachary I. and
         {Azalee Bostroem}, K. and {Burke}, D.~J. and {Casey}, Andrew R. and
         {Crawford}, Steven M. and {Dencheva}, Nadia and {Ely}, Justin and
         {Jenness}, Tim and {Labrie}, Kathleen and {Lim}, Pey Lian and
         {Pierfederici}, Francesco and {Pontzen}, Andrew and {Ptak}, Andy and
         {Refsdal}, Brian and {Servillat}, Mathieu and {Streicher}, Ole},
        title = "{Astropy: A community Python package for astronomy}",
      journal = {\aap},
         year = "2013",
        month = "Oct",
       volume = {558},
          eid = {A33},
        pages = {A33},
          doi = {10.1051/0004-6361/201322068},
archivePrefix = {arXiv},
       eprint = {1307.6212},
 primaryClass = {astro-ph.IM},
       adsurl = {https://ui.adsabs.harvard.edu/abs/2013A&A...558A..33A}
}

@Article{matplotlib2007,
  Author    = {Hunter, J. D.},
  Title     = {Matplotlib: A 2D graphics environment},
  Journal   = {Computing in Science \& Engineering},
  Volume    = {9},
  Number    = {3},
  Pages     = {90--95},
  publisher = {IEEE COMPUTER SOC},
  doi       = {10.1109/MCSE.2007.55},
  year      = 2007
}

@ARTICLE{scipy2020,
       author = {{Virtanen}, Pauli and {Gommers}, Ralf and {Oliphant},
         Travis E. and {Haberland}, Matt and {Reddy}, Tyler and
         {Cournapeau}, David and {Burovski}, Evgeni and {Peterson}, Pearu
         and {Weckesser}, Warren and {Bright}, Jonathan and {van der Walt},
         St{\'e}fan J.  and {Brett}, Matthew and {Wilson}, Joshua and
         {Jarrod Millman}, K.  and {Mayorov}, Nikolay and {Nelson}, Andrew
         R.~J. and {Jones}, Eric and {Kern}, Robert and {Larson}, Eric and
         {Carey}, CJ and {Polat}, {\.I}lhan and {Feng}, Yu and {Moore},
         Eric W. and {Vand erPlas}, Jake and {Laxalde}, Denis and
         {Perktold}, Josef and {Cimrman}, Robert and {Henriksen}, Ian and
         {Quintero}, E.~A. and {Harris}, Charles R and {Archibald}, Anne M.
         and {Ribeiro}, Ant{\^o}nio H. and {Pedregosa}, Fabian and
         {van Mulbregt}, Paul and {Contributors}, SciPy 1. 0},
        title = "{SciPy 1.0: Fundamental Algorithms for Scientific
                  Computing in Python}",
      journal = {Nature Methods},
      year = "2020",
      volume={17},
      pages={261--272},
      adsurl = {https://rdcu.be/b08Wh},
      doi = {https://doi.org/10.1038/s41592-019-0686-2},
}

@Article{harris2020array,
 title         = {Array programming with {NumPy}},
 author        = {Charles R. Harris and K. Jarrod Millman and St{\'{e}}fan J.
                 van der Walt and Ralf Gommers and Pauli Virtanen and David
                 Cournapeau and Eric Wieser and Julian Taylor and Sebastian
                 Berg and Nathaniel J. Smith and Robert Kern and Matti Picus
                 and Stephan Hoyer and Marten H. van Kerkwijk and Matthew
                 Brett and Allan Haldane and Jaime Fern{\'{a}}ndez del
                 R{\'{i}}o and Mark Wiebe and Pearu Peterson and Pierre
                 G{\'{e}}rard-Marchant and Kevin Sheppard and Tyler Reddy and
                 Warren Weckesser and Hameer Abbasi and Christoph Gohlke and
                 Travis E. Oliphant},
 year          = {2020},
 month         = sep,
 journal       = {Nature},
 volume        = {585},
 number        = {7825},
 pages         = {357--362},
 doi           = {10.1038/s41586-020-2649-2},
 publisher     = {Springer Science and Business Media {LLC}},
 url           = {https://doi.org/10.1038/s41586-020-2649-2}
}

@inproceedings{NEURIPS2019_bdbca288,
 author = {Paszke, Adam and Gross, Sam and Massa, Francisco and Lerer, Adam and Bradbury, James and Chanan, Gregory and Killeen, Trevor and Lin, Zeming and Gimelshein, Natalia and Antiga, Luca and Desmaison, Alban and Kopf, Andreas and Yang, Edward and DeVito, Zachary and Raison, Martin and Tejani, Alykhan and Chilamkurthy, Sasank and Steiner, Benoit and Fang, Lu and Bai, Junjie and Chintala, Soumith},
 booktitle = {Advances in Neural Information Processing Systems},
 editor = {H. Wallach and H. Larochelle and A. Beygelzimer and F. d\textquotesingle Alch\'{e}-Buc and E. Fox and R. Garnett},
 pages = {},
 publisher = {Curran Associates, Inc.},
 title = {PyTorch: An Imperative Style, High-Performance Deep Learning Library},
 url = {https://proceedings.neurips.cc/paper_files/paper/2019/file/bdbca288fee7f92f2bfa9f7012727740-Paper.pdf},
 volume = {32},
 year = {2019}
}

@ARTICLE{2022ApJ...931..127C,
       author = {{Cha}, Sangjun and {Jee}, M. James},
        title = "{MARS: A New Maximum-entropy-regularized Strong Lensing Mass Reconstruction Method}",
      journal = {\apj},
         year = 2022,
        month = jun,
       volume = {931},
       number = {2},
          eid = {127},
        pages = {127},
          doi = {10.3847/1538-4357/ac69df},
archivePrefix = {arXiv},
       eprint = {2202.10489},
 primaryClass = {astro-ph.CO},
       adsurl = {https://ui.adsabs.harvard.edu/abs/2022ApJ...931..127C}
}

@ARTICLE{2023ApJ...951..140C,
       author = {{Cha}, Sangjun and {Jee}, M. James},
        title = "{Model-independent Mass Reconstruction of the Hubble Frontier Field Clusters with MARS Based on Self-consistent Strong-lensing Data}",
      journal = {\apj},
         year = 2023,
        month = jul,
       volume = {951},
       number = {2},
          eid = {140},
        pages = {140},
          doi = {10.3847/1538-4357/acd111},
archivePrefix = {arXiv},
       eprint = {2301.08765},
 primaryClass = {astro-ph.GA},
       adsurl = {https://ui.adsabs.harvard.edu/abs/2023ApJ...951..140C}
}

@ARTICLE{2007ApJ...661..728J,
       author = {{Jee}, M.~J. and {Ford}, H.~C. and {Illingworth}, G.~D. and {White}, R.~L. and {Broadhurst}, T.~J. and {Coe}, D.~A. and {Meurer}, G.~R. and {van der Wel}, A. and {Ben{\'\i}tez}, N. and {Blakeslee}, J.~P. and {Bouwens}, R.~J. and {Bradley}, L.~D. and {Demarco}, R. and {Homeier}, N.~L. and {Martel}, A.~R. and {Mei}, S.},
        title = "{Discovery of a Ringlike Dark Matter Structure in the Core of the Galaxy Cluster Cl 0024+17}",
      journal = {\apj},
         year = 2007,
        month = jun,
       volume = {661},
       number = {2},
        pages = {728-749},
          doi = {10.1086/517498},
archivePrefix = {arXiv},
       eprint = {0705.2171},
 primaryClass = {astro-ph},
       adsurl = {https://ui.adsabs.harvard.edu/abs/2007ApJ...661..728J}
}

@Inbook{Kochanek2006,
author="Kochanek, C. S.",
title="Strong Gravitational Lensing",
bookTitle="Gravitational Lensing: Strong, Weak and Micro",
year="2006",
publisher="Springer Berlin Heidelberg",
address="Berlin, Heidelberg",
pages="91--268",
isbn="978-3-540-30310-7",
doi="10.1007/978-3-540-30310-7_2",
url="https://doi.org/10.1007/978-3-540-30310-7_2"
}

@ARTICLE{2011A&ARv..19...47K,
       author = {{Kneib}, Jean-Paul and {Natarajan}, Priyamvada},
        title = "{Cluster lenses}",
      journal = {\aapr},
         year = 2011,
        month = nov,
       volume = {19},
          eid = {47},
        pages = {47},
          doi = {10.1007/s00159-011-0047-3},
archivePrefix = {arXiv},
       eprint = {1202.0185},
 primaryClass = {astro-ph.CO},
       adsurl = {https://ui.adsabs.harvard.edu/abs/2011A&ARv..19...47K}
}

@ARTICLE{bartelmann2001,
       author = {{Bartelmann}, M. and {Schneider}, P.},
        title = "{Weak gravitational lensing}",
      journal = {\physrep},
         year = 2001,
        month = jan,
       volume = {340},
       number = {4-5},
        pages = {291-472},
          doi = {10.1016/S0370-1573(00)00082-X},
archivePrefix = {arXiv},
       eprint = {astro-ph/9912508},
 primaryClass = {astro-ph},
       adsurl = {https://ui.adsabs.harvard.edu/abs/2001PhR...340..291B}
}

@ARTICLE{hoekstra2013,
       author = {{Hoekstra}, Henk and {Bartelmann}, Matthias and {Dahle}, H{\r{a}}kon and {Israel}, Holger and {Limousin}, Marceau and {Meneghetti}, Massimo},
        title = "{Masses of Galaxy Clusters from Gravitational Lensing}",
      journal = {\ssr},
         year = 2013,
        month = aug,
       volume = {177},
       number = {1-4},
        pages = {75-118},
          doi = {10.1007/s11214-013-9978-5},
archivePrefix = {arXiv},
       eprint = {1303.3274},
 primaryClass = {astro-ph.CO},
       adsurl = {https://ui.adsabs.harvard.edu/abs/2013SSRv..177...75H}
}

@ARTICLE{2017ApJ...837...97L,
       author = {{Lotz}, J.~M. and {Koekemoer}, A. and {Coe}, D. and {Grogin}, N. and {Capak}, P. and {Mack}, J. and {Anderson}, J. and {Avila}, R. and {Barker}, E.~A. and {Borncamp}, D. and {Brammer}, G. and {Durbin}, M. and {Gunning}, H. and {Hilbert}, B. and {Jenkner}, H. and {Khandrika}, H. and {Levay}, Z. and {Lucas}, R.~A. and {MacKenty}, J. and {Ogaz}, S. and {Porterfield}, B. and {Reid}, N. and {Robberto}, M. and {Royle}, P. and {Smith}, L.~J. and {Storrie-Lombardi}, L.~J. and {Sunnquist}, B. and {Surace}, J. and {Taylor}, D.~C. and {Williams}, R. and {Bullock}, J. and {Dickinson}, M. and {Finkelstein}, S. and {Natarajan}, P. and {Richard}, J. and {Robertson}, B. and {Tumlinson}, J. and {Zitrin}, A. and {Flanagan}, K. and {Sembach}, K. and {Soifer}, B.~T. and {Mountain}, M.},
        title = "{The Frontier Fields: Survey Design and Initial Results}",
      journal = {\apj},
         year = 2017,
        month = mar,
       volume = {837},
       number = {1},
          eid = {97},
        pages = {97},
          doi = {10.3847/1538-4357/837/1/97},
archivePrefix = {arXiv},
       eprint = {1605.06567},
 primaryClass = {astro-ph.GA},
       adsurl = {https://ui.adsabs.harvard.edu/abs/2017ApJ...837...97L}
}

@ARTICLE{1998MNRAS.299..895B,
       author = {{Bridle}, S.~L. and {Hobson}, M.~P. and {Lasenby}, A.~N. and {Saunders}, Richard},
        title = "{A maximum-entropy method for reconstructing the projected mass distribution of gravitational lenses}",
      journal = {\mnras},
         year = 1998,
        month = sep,
       volume = {299},
       number = {3},
        pages = {895-903},
          doi = {10.1046/j.1365-8711.1998.01877.x},
archivePrefix = {arXiv},
       eprint = {astro-ph/9802159},
 primaryClass = {astro-ph},
       adsurl = {https://ui.adsabs.harvard.edu/abs/1998MNRAS.299..895B}
}

@ARTICLE{1996ApJ...462..563N,
       author = {{Navarro}, Julio F. and {Frenk}, Carlos S. and {White}, Simon D.~M.},
        title = "{The Structure of Cold Dark Matter Halos}",
      journal = {\apj},
         year = 1996,
        month = may,
       volume = {462},
        pages = {563},
          doi = {10.1086/177173},
archivePrefix = {arXiv},
       eprint = {astro-ph/9508025},
 primaryClass = {astro-ph},
       adsurl = {https://ui.adsabs.harvard.edu/abs/1996ApJ...462..563N}
}

@ARTICLE{1993ApJ...417..450K,
       author = {{Kassiola}, Aggeliki and {Kovner}, Israel},
        title = "{Elliptic Mass Distributions versus Elliptic Potentials in Gravitational Lenses}",
      journal = {\apj},
         year = 1993,
        month = nov,
       volume = {417},
        pages = {450},
          doi = {10.1086/173325},
       adsurl = {https://ui.adsabs.harvard.edu/abs/1993ApJ...417..450K}
}

@ARTICLE{2017MNRAS.472.3177M,
       author = {{Meneghetti}, M. and {Natarajan}, P. and {Coe}, D. and {Contini}, E. and {De Lucia}, G. and {Giocoli}, C. and {Acebron}, A. and {Borgani}, S. and {Bradac}, M. and {Diego}, J.~M. and {Hoag}, A. and {Ishigaki}, M. and {Johnson}, T.~L. and {Jullo}, E. and {Kawamata}, R. and {Lam}, D. and {Limousin}, M. and {Liesenborgs}, J. and {Oguri}, M. and {Sebesta}, K. and {Sharon}, K. and {Williams}, L.~L.~R. and {Zitrin}, A.},
        title = "{The Frontier Fields lens modelling comparison project}",
      journal = {\mnras},
         year = 2017,
        month = dec,
       volume = {472},
       number = {3},
        pages = {3177-3216},
          doi = {10.1093/mnras/stx2064},
archivePrefix = {arXiv},
       eprint = {1606.04548},
 primaryClass = {astro-ph.CO},
       adsurl = {https://ui.adsabs.harvard.edu/abs/2017MNRAS.472.3177M}
}

@ARTICLE{2018AJ....156..123A,
       author = {{Astropy Collaboration} and {Price-Whelan}, A.~M. and {Sip{\H{o}}cz}, B.~M. and {G{\"u}nther}, H.~M. and {Lim}, P.~L. and {Crawford}, S.~M. and {Conseil}, S. and {Shupe}, D.~L. and {Craig}, M.~W. and {Dencheva}, N. and {Ginsburg}, A. and {VanderPlas}, J.~T. and {Bradley}, L.~D. and {P{\'e}rez-Su{\'a}rez}, D. and {de Val-Borro}, M. and {Aldcroft}, T.~L. and {Cruz}, K.~L. and {Robitaille}, T.~P. and {Tollerud}, E.~J. and {Ardelean}, C. and {Babej}, T. and {Bach}, Y.~P. and {Bachetti}, M. and {Bakanov}, A.~V. and {Bamford}, S.~P. and {Barentsen}, G. and {Barmby}, P. and {Baumbach}, A. and {Berry}, K.~L. and {Biscani}, F. and {Boquien}, M. and {Bostroem}, K.~A. and {Bouma}, L.~G. and {Brammer}, G.~B. and {Bray}, E.~M. and {Breytenbach}, H. and {Buddelmeijer}, H. and {Burke}, D.~J. and {Calderone}, G. and {Cano Rodr{\'\i}guez}, J.~L. and {Cara}, M. and {Cardoso}, J.~V.~M. and {Cheedella}, S. and {Copin}, Y. and {Corrales}, L. and {Crichton}, D. and {D'Avella}, D. and {Deil}, C. and {Depagne}, {\'E}. and {Dietrich}, J.~P. and {Donath}, A. and {Droettboom}, M. and {Earl}, N. and {Erben}, T. and {Fabbro}, S. and {Ferreira}, L.~A. and {Finethy}, T. and {Fox}, R.~T. and {Garrison}, L.~H. and {Gibbons}, S.~L.~J. and {Goldstein}, D.~A. and {Gommers}, R. and {Greco}, J.~P. and {Greenfield}, P. and {Groener}, A.~M. and {Grollier}, F. and {Hagen}, A. and {Hirst}, P. and {Homeier}, D. and {Horton}, A.~J. and {Hosseinzadeh}, G. and {Hu}, L. and {Hunkeler}, J.~S. and {Ivezi{\'c}}, {\v{Z}}. and {Jain}, A. and {Jenness}, T. and {Kanarek}, G. and {Kendrew}, S. and {Kern}, N.~S. and {Kerzendorf}, W.~E. and {Khvalko}, A. and {King}, J. and {Kirkby}, D. and {Kulkarni}, A.~M. and {Kumar}, A. and {Lee}, A. and {Lenz}, D. and {Littlefair}, S.~P. and {Ma}, Z. and {Macleod}, D.~M. and {Mastropietro}, M. and {McCully}, C. and {Montagnac}, S. and {Morris}, B.~M. and {Mueller}, M. and {Mumford}, S.~J. and {Muna}, D. and {Murphy}, N.~A. and {Nelson}, S. and {Nguyen}, G.~H. and {Ninan}, J.~P. and {N{\"o}the}, M. and {Ogaz}, S. and {Oh}, S. and {Parejko}, J.~K. and {Parley}, N. and {Pascual}, S. and {Patil}, R. and {Patil}, A.~A. and {Plunkett}, A.~L. and {Prochaska}, J.~X. and {Rastogi}, T. and {Reddy Janga}, V. and {Sabater}, J. and {Sakurikar}, P. and {Seifert}, M. and {Sherbert}, L.~E. and {Sherwood-Taylor}, H. and {Shih}, A.~Y. and {Sick}, J. and {Silbiger}, M.~T. and {Singanamalla}, S. and {Singer}, L.~P. and {Sladen}, P.~H. and {Sooley}, K.~A. and {Sornarajah}, S. and {Streicher}, O. and {Teuben}, P. and {Thomas}, S.~W. and {Tremblay}, G.~R. and {Turner}, J.~E.~H. and {Terr{\'o}n}, V. and {van Kerkwijk}, M.~H. and {de la Vega}, A. and {Watkins}, L.~L. and {Weaver}, B.~A. and {Whitmore}, J.~B. and {Woillez}, J. and {Zabalza}, V. and {Astropy Contributors}},
        title = "{The Astropy Project: Building an Open-science Project and Status of the v2.0 Core Package}",
      journal = {\aj},
         year = 2018,
        month = sep,
       volume = {156},
       number = {3},
          eid = {123},
        pages = {123},
          doi = {10.3847/1538-3881/aabc4f},
archivePrefix = {arXiv},
       eprint = {1801.02634},
 primaryClass = {astro-ph.IM},
       adsurl = {https://ui.adsabs.harvard.edu/abs/2018AJ....156..123A}
}

@ARTICLE{2022ApJ...935..167A,
       author = {{Astropy Collaboration} and {Price-Whelan}, Adrian M. and {Lim}, Pey Lian and {Earl}, Nicholas and {Starkman}, Nathaniel and {Bradley}, Larry and {Shupe}, David L. and {Patil}, Aarya A. and {Corrales}, Lia and {Brasseur}, C.~E. and {N{\"o}the}, Maximilian and {Donath}, Axel and {Tollerud}, Erik and {Morris}, Brett M. and {Ginsburg}, Adam and {Vaher}, Eero and {Weaver}, Benjamin A. and {Tocknell}, James and {Jamieson}, William and {van Kerkwijk}, Marten H. and {Robitaille}, Thomas P. and {Merry}, Bruce and {Bachetti}, Matteo and {G{\"u}nther}, H. Moritz and {Aldcroft}, Thomas L. and {Alvarado-Montes}, Jaime A. and {Archibald}, Anne M. and {B{\'o}di}, Attila and {Bapat}, Shreyas and {Barentsen}, Geert and {Baz{\'a}n}, Juanjo and {Biswas}, Manish and {Boquien}, M{\'e}d{\'e}ric and {Burke}, D.~J. and {Cara}, Daria and {Cara}, Mihai and {Conroy}, Kyle E. and {Conseil}, Simon and {Craig}, Matthew W. and {Cross}, Robert M. and {Cruz}, Kelle L. and {D'Eugenio}, Francesco and {Dencheva}, Nadia and {Devillepoix}, Hadrien A.~R. and {Dietrich}, J{\"o}rg P. and {Eigenbrot}, Arthur Davis and {Erben}, Thomas and {Ferreira}, Leonardo and {Foreman-Mackey}, Daniel and {Fox}, Ryan and {Freij}, Nabil and {Garg}, Suyog and {Geda}, Robel and {Glattly}, Lauren and {Gondhalekar}, Yash and {Gordon}, Karl D. and {Grant}, David and {Greenfield}, Perry and {Groener}, Austen M. and {Guest}, Steve and {Gurovich}, Sebastian and {Handberg}, Rasmus and {Hart}, Akeem and {Hatfield-Dodds}, Zac and {Homeier}, Derek and {Hosseinzadeh}, Griffin and {Jenness}, Tim and {Jones}, Craig K. and {Joseph}, Prajwel and {Kalmbach}, J. Bryce and {Karamehmetoglu}, Emir and {Ka{\l}uszy{\'n}ski}, Miko{\l}aj and {Kelley}, Michael S.~P. and {Kern}, Nicholas and {Kerzendorf}, Wolfgang E. and {Koch}, Eric W. and {Kulumani}, Shankar and {Lee}, Antony and {Ly}, Chun and {Ma}, Zhiyuan and {MacBride}, Conor and {Maljaars}, Jakob M. and {Muna}, Demitri and {Murphy}, N.~A. and {Norman}, Henrik and {O'Steen}, Richard and {Oman}, Kyle A. and {Pacifici}, Camilla and {Pascual}, Sergio and {Pascual-Granado}, J. and {Patil}, Rohit R. and {Perren}, Gabriel I. and {Pickering}, Timothy E. and {Rastogi}, Tanuj and {Roulston}, Benjamin R. and {Ryan}, Daniel F. and {Rykoff}, Eli S. and {Sabater}, Jose and {Sakurikar}, Parikshit and {Salgado}, Jes{\'u}s and {Sanghi}, Aniket and {Saunders}, Nicholas and {Savchenko}, Volodymyr and {Schwardt}, Ludwig and {Seifert-Eckert}, Michael and {Shih}, Albert Y. and {Jain}, Anany Shrey and {Shukla}, Gyanendra and {Sick}, Jonathan and {Simpson}, Chris and {Singanamalla}, Sudheesh and {Singer}, Leo P. and {Singhal}, Jaladh and {Sinha}, Manodeep and {Sip{\H{o}}cz}, Brigitta M. and {Spitler}, Lee R. and {Stansby}, David and {Streicher}, Ole and {{\v{S}}umak}, Jani and {Swinbank}, John D. and {Taranu}, Dan S. and {Tewary}, Nikita and {Tremblay}, Grant R. and {de Val-Borro}, Miguel and {Van Kooten}, Samuel J. and {Vasovi{\'c}}, Zlatan and {Verma}, Shresth and {de Miranda Cardoso}, Jos{\'e} Vin{\'\i}cius and {Williams}, Peter K.~G. and {Wilson}, Tom J. and {Winkel}, Benjamin and {Wood-Vasey}, W.~M. and {Xue}, Rui and {Yoachim}, Peter and {Zhang}, Chen and {Zonca}, Andrea and {Astropy Project Contributors}},
        title = "{The Astropy Project: Sustaining and Growing a Community-oriented Open-source Project and the Latest Major Release (v5.0) of the Core Package}",
      journal = {\apj},
         year = 2022,
        month = aug,
       volume = {935},
       number = {2},
          eid = {167},
        pages = {167},
          doi = {10.3847/1538-4357/ac7c74},
archivePrefix = {arXiv},
       eprint = {2206.14220},
 primaryClass = {astro-ph.IM},
       adsurl = {https://ui.adsabs.harvard.edu/abs/2022ApJ...935..167A}
}

@ARTICLE{2025ApJ...979...13P,
       author = {{Pascale}, Massimo and {Frye}, Brenda L. and {Pierel}, Justin D.~R. and {Chen}, Wenlei and {Kelly}, Patrick L. and {Cohen}, Seth H. and {Windhorst}, Rogier A. and {Riess}, Adam G. and {Kamieneski}, Patrick S. and {Diego}, Jos{\'e} M. and {Meena}, Ashish K. and {Cha}, Sangjun and {Oguri}, Masamune and {Zitrin}, Adi and {Jee}, M. James and {Foo}, Nicholas and {Leimbach}, Reagen and {Koekemoer}, Anton M. and {Conselice}, C.~J. and {Dai}, Liang and {Goobar}, Ariel and {Siebert}, Matthew R. and {Strolger}, Lou and {Willner}, S.~P.},
        title = "{SN H0pe: The First Measurement of H$_{0}$ from a Multiply Imaged Type Ia Supernova, Discovered by JWST}",
      journal = {\apj},
         year = 2025,
        month = jan,
       volume = {979},
       number = {1},
          eid = {13},
        pages = {13},
          doi = {10.3847/1538-4357/ad9928},
archivePrefix = {arXiv},
       eprint = {2403.18902},
 primaryClass = {astro-ph.CO},
       adsurl = {https://ui.adsabs.harvard.edu/abs/2025ApJ...979...13P}
}

@ARTICLE{2024ApJ...961..186C,
       author = {{Cha}, Sangjun and {HyeongHan}, Kim and {Scofield}, Zachary P. and {Joo}, Hyungjin and {Jee}, M. James},
        title = "{Precision MARS Mass Reconstruction of A2744: Synergizing the Largest Strong-lensing and Densest Weak-lensing Data Sets from JWST}",
      journal = {\apj},
         year = 2024,
        month = feb,
       volume = {961},
       number = {2},
          eid = {186},
        pages = {186},
          doi = {10.3847/1538-4357/ad0cbf},
archivePrefix = {arXiv},
       eprint = {2308.14805},
 primaryClass = {astro-ph.GA},
       adsurl = {https://ui.adsabs.harvard.edu/abs/2024ApJ...961..186C}
}

@ARTICLE{2002A&A...390..821G,
       author = {{Golse}, G. and {Kneib}, J. -P.},
        title = "{Pseudo elliptical lensing mass model: Application to the NFW mass distribution}",
      journal = {\aap},
         year = 2002,
        month = aug,
       volume = {390},
        pages = {821-827},
          doi = {10.1051/0004-6361:20020639},
archivePrefix = {arXiv},
       eprint = {astro-ph/0112138},
 primaryClass = {astro-ph},
       adsurl = {https://ui.adsabs.harvard.edu/abs/2002A&A...390..821G}
}

@ARTICLE{2012A&A...544A..83D,
       author = {{D{\'u}met-Montoya}, H.~S. and {Caminha}, G.~B. and {Makler}, M.},
        title = "{Domain of validity for pseudo-elliptical NFW lens models. Mass distribution, mapping to elliptical models, and arc cross section}",
      journal = {\aap},
         year = 2012,
        month = aug,
       volume = {544},
          eid = {A83},
        pages = {A83},
          doi = {10.1051/0004-6361/201118485},
archivePrefix = {arXiv},
       eprint = {1208.5682},
 primaryClass = {astro-ph.CO},
       adsurl = {https://ui.adsabs.harvard.edu/abs/2012A&A...544A..83D}
}

@ARTICLE{2009JCAP...01..015B,
       author = {{Baltz}, Edward A. and {Marshall}, Phil and {Oguri}, Masamune},
        title = "{Analytic models of plausible gravitational lens potentials}",
      journal = {\jcap},
         year = 2009,
        month = jan,
       volume = {2009},
       number = {1},
          eid = {015},
        pages = {015},
          doi = {10.1088/1475-7516/2009/01/015},
archivePrefix = {arXiv},
       eprint = {0705.0682},
 primaryClass = {astro-ph},
       adsurl = {https://ui.adsabs.harvard.edu/abs/2009JCAP...01..015B}
}

@ARTICLE{2024SSRv..220...19N,
       author = {{Natarajan}, P. and {Williams}, L.~L.~R. and {Brada{\v{c}}}, M. and {Grillo}, C. and {Ghosh}, A. and {Sharon}, K. and {Wagner}, J.},
        title = "{Strong Lensing by Galaxy Clusters}",
      journal = {\ssr},
         year = 2024,
        month = feb,
       volume = {220},
       number = {2},
          eid = {19},
        pages = {19},
          doi = {10.1007/s11214-024-01051-8},
archivePrefix = {arXiv},
       eprint = {2403.06245},
 primaryClass = {astro-ph.CO},
       adsurl = {https://ui.adsabs.harvard.edu/abs/2024SSRv..220...19N}
}

@ARTICLE{1998MNRAS.298..905H,
       author = {{Hobson}, M.~P. and {Lasenby}, A.~N.},
        title = "{The entropic prior for distributions with positive and negative values}",
      journal = {\mnras},
         year = 1998,
        month = aug,
       volume = {298},
       number = {3},
        pages = {905-908},
          doi = {10.1046/j.1365-8711.1998.01707.x},
archivePrefix = {arXiv},
       eprint = {astro-ph/9810240},
 primaryClass = {astro-ph},
       adsurl = {https://ui.adsabs.harvard.edu/abs/1998MNRAS.298..905H}
}

@ARTICLE{2019MNRAS.482.5666W,
       author = {{Williams}, Liliya L.~R. and {Liesenborgs}, Jori},
        title = "{The role of multiple images and model priors in measuring H$_{0}$ from supernova Refsdal in galaxy cluster MACS J1149.5+2223}",
      journal = {\mnras},
         year = 2019,
        month = feb,
       volume = {482},
       number = {4},
        pages = {5666-5677},
          doi = {10.1093/mnras/sty3113},
archivePrefix = {arXiv},
       eprint = {1806.11113},
 primaryClass = {astro-ph.CO},
       adsurl = {https://ui.adsabs.harvard.edu/abs/2019MNRAS.482.5666W}
}

@ARTICLE{1987ApJ...321..658B,
       author = {{Blandford}, Roger D. and {Kochanek}, Christopher S.},
        title = "{Gravitational Imaging by Isolated Elliptical Potential Wells. I. Cross Sections}",
      journal = {\apj},
         year = 1987,
        month = oct,
       volume = {321},
        pages = {658},
          doi = {10.1086/165660},
       adsurl = {https://ui.adsabs.harvard.edu/abs/1987ApJ...321..658B}
}

@ARTICLE{1996A&A...313..697B,
       author = {{Bartelmann}, M.},
        title = "{Arcs from a universal dark-matter halo profile.}",
      journal = {\aap},
         year = 1996,
        month = sep,
       volume = {313},
        pages = {697-702},
          doi = {10.48550/arXiv.astro-ph/9602053},
archivePrefix = {arXiv},
       eprint = {astro-ph/9602053},
 primaryClass = {astro-ph},
       adsurl = {https://ui.adsabs.harvard.edu/abs/1996A&A...313..697B}
}

@ARTICLE{2007A&A...461..881L,
       author = {{Limousin}, M. and {Kneib}, J.~P. and {Bardeau}, S. and {Natarajan}, P. and {Czoske}, O. and {Smail}, I. and {Ebeling}, H. and {Smith}, G.~P.},
        title = "{Truncation of galaxy dark matter halos in high density environments}",
      journal = {\aap},
         year = 2007,
        month = jan,
       volume = {461},
       number = {3},
        pages = {881-891},
          doi = {10.1051/0004-6361:20065543},
archivePrefix = {arXiv},
       eprint = {astro-ph/0609782},
 primaryClass = {astro-ph},
       adsurl = {https://ui.adsabs.harvard.edu/abs/2007A&A...461..881L}
}

@ARTICLE{2009ApJ...696.1771L,
       author = {{Limousin}, Marceau and {Sommer-Larsen}, Jesper and {Natarajan}, Priyamvada and {Milvang-Jensen}, Bo},
        title = "{Probing the Truncation of Galaxy Dark Matter Halos in High-Density Environments from Hydrodynamical N-Body Simulations}",
      journal = {\apj},
         year = 2009,
        month = may,
       volume = {696},
       number = {2},
        pages = {1771-1779},
          doi = {10.1088/0004-637X/696/2/1771},
archivePrefix = {arXiv},
       eprint = {0706.3149},
 primaryClass = {astro-ph},
       adsurl = {https://ui.adsabs.harvard.edu/abs/2009ApJ...696.1771L}
}

@ARTICLE{2017ApJ...841...18B,
       author = {{Baxter}, Eric and {Chang}, Chihway and {Jain}, Bhuvnesh and {Adhikari}, Susmita and {Dalal}, Neal and {Kravtsov}, Andrey and {More}, Surhud and {Rozo}, Eduardo and {Rykoff}, Eli and {Sheth}, Ravi K.},
        title = "{The Halo Boundary of Galaxy Clusters in the SDSS}",
      journal = {\apj},
         year = 2017,
        month = may,
       volume = {841},
       number = {1},
          eid = {18},
        pages = {18},
          doi = {10.3847/1538-4357/aa6ff0},
archivePrefix = {arXiv},
       eprint = {1702.01722},
 primaryClass = {astro-ph.CO},
       adsurl = {https://ui.adsabs.harvard.edu/abs/2017ApJ...841...18B}
}

@ARTICLE{2011MNRAS.414.1851O,
       author = {{Oguri}, Masamune and {Hamana}, Takashi},
        title = "{Detailed cluster lensing profiles at large radii and the impact on cluster weak lensing studies}",
      journal = {\mnras},
         year = 2011,
        month = jul,
       volume = {414},
       number = {3},
        pages = {1851-1861},
          doi = {10.1111/j.1365-2966.2011.18481.x},
archivePrefix = {arXiv},
       eprint = {1101.0650},
 primaryClass = {astro-ph.CO},
       adsurl = {https://ui.adsabs.harvard.edu/abs/2011MNRAS.414.1851O}
}

@ARTICLE{2023A&A...679A..31D,
       author = {{Diego}, Jose M. and {Sun}, Bangzheng and {Yan}, Haojing and {Furtak}, Lukas J. and {Zackrisson}, Erik and {Dai}, Liang and {Kelly}, Patrick and {Nonino}, Mario and {Adams}, Nathan and {Meena}, Ashish K. and {Willner}, Steven P. and {Zitrin}, Adi and {Cohen}, Seth H. and {D'Silva}, Jordan C.~J. and {Jansen}, Rolf A. and {Summers}, Jake and {Windhorst}, Rogier A. and {Coe}, Dan and {Conselice}, Christopher J. and {Driver}, Simon P. and {Frye}, Brenda and {Grogin}, Norman A. and {Koekemoer}, Anton M. and {Marshall}, Madeline A. and {Pirzkal}, Nor and {Robotham}, Aaron and {Rutkowski}, Michael J. and {Ryan}, Russell E. and {Tompkins}, Scott and {Willmer}, Christopher N.~A. and {Bhatawdekar}, Rachana},
        title = "{JWST's PEARLS: Mothra, a new kaiju star at z = 2.091 extremely magnified by MACS0416, and implications for dark matter models}",
      journal = {\aap},
         year = 2023,
        month = nov,
       volume = {679},
          eid = {A31},
        pages = {A31},
          doi = {10.1051/0004-6361/202347556},
archivePrefix = {arXiv},
       eprint = {2307.10363},
 primaryClass = {astro-ph.CO},
       adsurl = {https://ui.adsabs.harvard.edu/abs/2023A&A...679A..31D}
}

@ARTICLE{2025A&A...696A..15R,
       author = {{Rihtar{\v{s}}i{\v{c}}}, G. and {Brada{\v{c}}}, M. and {Desprez}, G. and {Harshan}, A. and {Noirot}, G. and {Estrada-Carpenter}, V. and {Martis}, N.~S. and {Abraham}, R.~G. and {Asada}, Y. and {Brammer}, G. and {Iyer}, K.~G. and {Matharu}, J. and {Mowla}, L. and {Muzzin}, A. and {Sarrouh}, G.~T.~E. and {Sawicki}, M. and {Strait}, V. and {Willott}, C.~J. and {Gledhill}, R. and {Markov}, V. and {Tripodi}, R.},
        title = "{CANUCS: Constraining the MACS J0416.1-2403 strong lensing model with JWST NIRISS, NIRSpec, and NIRCam}",
      journal = {\aap},
         year = 2025,
        month = apr,
       volume = {696},
          eid = {A15},
        pages = {A15},
          doi = {10.1051/0004-6361/202451117},
       adsurl = {https://ui.adsabs.harvard.edu/abs/2025A&A...696A..15R}
}

@ARTICLE{2025MNRAS.536.2690P,
       author = {{Perera}, Derek and {Williams}, Liliya L.~R. and {Liesenborgs}, Jori and {Kelly}, Patrick L. and {Taft}, Sarah H. and {Li}, Sung Kei and {Jauzac}, Mathilde and {Diego}, Jose M. and {Natarajan}, Priyamvada and {Steinhardt}, Charles L. and {Faisst}, Andreas L. and {Rich}, R. Michael and {Limousin}, Marceau},
        title = "{BUFFALO wild wings: a high-precision free-form lens model of MACSJ0416 with constraints on dark matter from substructure and highly magnified arcs}",
      journal = {\mnras},
         year = 2025,
        month = jan,
       volume = {536},
       number = {3},
        pages = {2690-2713},
          doi = {10.1093/mnras/stae2753},
archivePrefix = {arXiv},
       eprint = {2407.15978},
 primaryClass = {astro-ph.GA},
       adsurl = {https://ui.adsabs.harvard.edu/abs/2025MNRAS.536.2690P}
}

@ARTICLE{2023A&A...674A..79B,
       author = {{Bergamini}, P. and {Grillo}, C. and {Rosati}, P. and {Vanzella}, E. and {Me{\v{s}}tri{\'c}}, U. and {Mercurio}, A. and {Acebron}, A. and {Caminha}, G.~B. and {Granata}, G. and {Meneghetti}, M. and {Angora}, G. and {Nonino}, M.},
        title = "{A state-of-the-art strong-lensing model of MACS J0416.1{\ensuremath{-}}2403 with the largest sample of spectroscopic multiple images}",
      journal = {\aap},
         year = 2023,
        month = jun,
       volume = {674},
          eid = {A79},
        pages = {A79},
          doi = {10.1051/0004-6361/202244834},
archivePrefix = {arXiv},
       eprint = {2208.14020},
 primaryClass = {astro-ph.CO},
       adsurl = {https://ui.adsabs.harvard.edu/abs/2023A&A...674A..79B}
}

@ARTICLE{2021A&A...646A..83R,
       author = {{Richard}, Johan and {Claeyssens}, Ad{\'e}la{\"\i}de and {Lagattuta}, David and {Guaita}, Lucia and {Bauer}, Franz Erik and {Pello}, Roser and {Carton}, David and {Bacon}, Roland and {Soucail}, Genevi{\`e}ve and {Lyon}, Gonzalo Prieto and {Kneib}, Jean-Paul and {Mahler}, Guillaume and {Cl{\'e}ment}, Benjamin and {Mercier}, Wilfried and {Variu}, Andrei and {Tamone}, Am{\'e}lie and {Ebeling}, Harald and {Schmidt}, Kasper B. and {Nanayakkara}, Themiya and {Maseda}, Michael and {Weilbacher}, Peter M. and {Bouch{\'e}}, Nicolas and {Bouwens}, Rychard J. and {Wisotzki}, Lutz and {de la Vieuville}, Geoffroy and {Martinez}, Johany and {Patr{\'\i}cio}, Vera},
        title = "{An atlas of MUSE observations towards twelve massive lensing clusters}",
      journal = {\aap},
         year = 2021,
        month = feb,
       volume = {646},
          eid = {A83},
        pages = {A83},
          doi = {10.1051/0004-6361/202039462},
archivePrefix = {arXiv},
       eprint = {2009.09784},
 primaryClass = {astro-ph.GA},
       adsurl = {https://ui.adsabs.harvard.edu/abs/2021A&A...646A..83R}
}

@ARTICLE{2024A&A...681A.124D,
       author = {{Diego}, Jose M. and {Li}, Sung Kei and {Meena}, Ashish K. and {Niemiec}, Anna and {Acebron}, Ana and {Jauzac}, Mathilde and {Struble}, Mitchell F. and {Amruth}, Alfred and {Broadhurst}, Tom J. and {Cerny}, Catherine and {Ebeling}, Harald and {Filippenko}, Alexei V. and {Jullo}, Eric and {Kelly}, Patrick and {Koekemoer}, Anton M. and {Lagattuta}, David and {Lim}, Jeremy and {Limousin}, Marceau and {Mahler}, Guillaume and {Patel}, Nency and {Remolina}, Juan and {Richard}, Johan and {Sharon}, Keren and {Steinhardt}, Charles and {Umetsu}, Keiichi and {Williams}, Liliya and {Zitrin}, Adi and {Palencia}, Jose Mar{\'\i}a and {Dai}, Liang and {Ji}, Lingyuan and {Pascale}, Massimo},
        title = "{BUFFALO/Flashlights: Constraints on the abundance of lensed supergiant stars in the Spock galaxy at redshift 1}",
      journal = {\aap},
         year = 2024,
        month = jan,
       volume = {681},
          eid = {A124},
        pages = {A124},
          doi = {10.1051/0004-6361/202346761},
archivePrefix = {arXiv},
       eprint = {2304.09222},
 primaryClass = {astro-ph.GA},
       adsurl = {https://ui.adsabs.harvard.edu/abs/2024A&A...681A.124D}
}

@ARTICLE{2007NJPh....9..447J,
       author = {{Jullo}, E. and {Kneib}, J. -P. and {Limousin}, M. and {El{\'\i}asd{\'o}ttir}, {\'A}. and {Marshall}, P.~J. and {Verdugo}, T.},
        title = "{A Bayesian approach to strong lensing modelling of galaxy clusters}",
      journal = {New Journal of Physics},
         year = 2007,
        month = dec,
       volume = {9},
       number = {12},
        pages = {447},
          doi = {10.1088/1367-2630/9/12/447},
archivePrefix = {arXiv},
       eprint = {0706.0048},
 primaryClass = {astro-ph},
       adsurl = {https://ui.adsabs.harvard.edu/abs/2007NJPh....9..447J}
}

@ARTICLE{2009MNRAS.395.1319J,
       author = {{Jullo}, E. and {Kneib}, J. -P.},
        title = "{Multiscale cluster lens mass mapping - I. Strong lensing modelling}",
      journal = {\mnras},
         year = 2009,
        month = may,
       volume = {395},
       number = {3},
        pages = {1319-1332},
          doi = {10.1111/j.1365-2966.2009.14654.x},
archivePrefix = {arXiv},
       eprint = {0901.3792},
 primaryClass = {astro-ph.CO},
       adsurl = {https://ui.adsabs.harvard.edu/abs/2009MNRAS.395.1319J}
}

@ARTICLE{2017ApJ...851...46F,
       author = {{Finner}, Kyle and {Jee}, M. James and {Golovich}, Nathan and {Wittman}, David and {Dawson}, William and {Gruen}, Daniel and {Koekemoer}, Anton M. and {Lemaux}, Brian C. and {Seitz}, Stella},
        title = "{MC$^{2}$: Subaru and Hubble Space Telescope Weak-lensing Analysis of the Double Radio Relic Galaxy Cluster PLCK G287.0+32.9}",
      journal = {\apj},
         year = 2017,
        month = dec,
       volume = {851},
       number = {1},
          eid = {46},
        pages = {46},
          doi = {10.3847/1538-4357/aa998c},
archivePrefix = {arXiv},
       eprint = {1710.02527},
 primaryClass = {astro-ph.CO},
       adsurl = {https://ui.adsabs.harvard.edu/abs/2017ApJ...851...46F}
}

@ARTICLE{2019A&A...631A.130B,
       author = {{Bergamini}, P. and {Rosati}, P. and {Mercurio}, A. and {Grillo}, C. and {Caminha}, G.~B. and {Meneghetti}, M. and {Agnello}, A. and {Biviano}, A. and {Calura}, F. and {Giocoli}, C. and {Lombardi}, M. and {Rodighiero}, G. and {Vanzella}, E.},
        title = "{Enhanced cluster lensing models with measured galaxy kinematics}",
      journal = {\aap},
         year = 2019,
        month = nov,
       volume = {631},
          eid = {A130},
        pages = {A130},
          doi = {10.1051/0004-6361/201935974},
archivePrefix = {arXiv},
       eprint = {1905.13236},
 primaryClass = {astro-ph.GA},
       adsurl = {https://ui.adsabs.harvard.edu/abs/2019A&A...631A.130B}
}

@ARTICLE{2021ApJ...923..101K,
       author = {{Kim}, Jinhyub and {Jee}, M. James and {Hughes}, John P. and {Yoon}, Mijin and {HyeongHan}, Kim and {Menanteau}, Felipe and {Sif{\'o}n}, Crist{\'o}bal and {Hovey}, Luke and {Arunachalam}, Prasiddha},
        title = "{Head-to-Toe Measurement of El Gordo: Improved Analysis of the Galaxy Cluster ACT-CL J0102-4915 with New Wide-field Hubble Space Telescope Imaging Data}",
      journal = {\apj},
         year = 2021,
        month = dec,
       volume = {923},
       number = {1},
          eid = {101},
        pages = {101},
          doi = {10.3847/1538-4357/ac294f},
archivePrefix = {arXiv},
       eprint = {2106.00031},
 primaryClass = {astro-ph.CO},
       adsurl = {https://ui.adsabs.harvard.edu/abs/2021ApJ...923..101K}
}

@ARTICLE{2022A&A...657A..83C,
       author = {{Caminha}, G.~B. and {Suyu}, S.~H. and {Grillo}, C. and {Rosati}, P.},
        title = "{Galaxy cluster strong lensing cosmography. Cosmological constraints from a sample of regular galaxy clusters}",
      journal = {\aap},
         year = 2022,
        month = jan,
       volume = {657},
          eid = {A83},
        pages = {A83},
          doi = {10.1051/0004-6361/202141994},
archivePrefix = {arXiv},
       eprint = {2110.06232},
 primaryClass = {astro-ph.CO},
       adsurl = {https://ui.adsabs.harvard.edu/abs/2022A&A...657A..83C}
}

@ARTICLE{2017MNRAS.470.1809A,
       author = {{Acebron}, Ana and {Jullo}, Eric and {Limousin}, Marceau and {Tilquin}, Andr{\'e} and {Giocoli}, Carlo and {Jauzac}, Mathilde and {Mahler}, Guillaume and {Richard}, Johan},
        title = "{Hubble Frontier Fields: systematic errors in strong lensing models of galaxy clusters - implications for cosmography}",
      journal = {\mnras},
         year = 2017,
        month = sep,
       volume = {470},
       number = {2},
        pages = {1809-1825},
          doi = {10.1093/mnras/stx1330},
archivePrefix = {arXiv},
       eprint = {1704.05380},
 primaryClass = {astro-ph.CO},
       adsurl = {https://ui.adsabs.harvard.edu/abs/2017MNRAS.470.1809A}
}

@ARTICLE{2018ApJ...865..122M,
       author = {{Maga{\~n}a}, Juan and {Acebr{\'o}n}, Ana and {Motta}, Ver{\'o}nica and {Verdugo}, Tom{\'a}s and {Jullo}, Eric and {Limousin}, Marceau},
        title = "{Strong Lensing Modeling in Galaxy Clusters as a Promising Method to Test Cosmography. I. Parametric Dark Energy Models}",
      journal = {\apj},
         year = 2018,
        month = oct,
       volume = {865},
       number = {2},
          eid = {122},
        pages = {122},
          doi = {10.3847/1538-4357/aada7d},
archivePrefix = {arXiv},
       eprint = {1711.00829},
 primaryClass = {astro-ph.CO},
       adsurl = {https://ui.adsabs.harvard.edu/abs/2018ApJ...865..122M}
}

@ARTICLE{2024A&A...687A.270R,
       author = {{Ragagnin}, A. and {Meneghetti}, M. and {Calura}, F. and {Despali}, G. and {Dolag}, K. and {Fischer}, M.~S. and {Giocoli}, C. and {Moscardini}, L.},
        title = "{Dianoga SIDM: Galaxy cluster self-interacting dark matter simulations}",
      journal = {\aap},
         year = 2024,
        month = jul,
       volume = {687},
          eid = {A270},
        pages = {A270},
          doi = {10.1051/0004-6361/202449872},
archivePrefix = {arXiv},
       eprint = {2404.01383},
 primaryClass = {astro-ph.CO},
       adsurl = {https://ui.adsabs.harvard.edu/abs/2024A&A...687A.270R}
}

@ARTICLE{2019MNRAS.488.3646R,
       author = {{Robertson}, Andrew and {Harvey}, David and {Massey}, Richard and {Eke}, Vincent and {McCarthy}, Ian G. and {Jauzac}, Mathilde and {Li}, Baojiu and {Schaye}, Joop},
        title = "{Observable tests of self-interacting dark matter in galaxy clusters: cosmological simulations with SIDM and baryons}",
      journal = {\mnras},
         year = 2019,
        month = sep,
       volume = {488},
       number = {3},
        pages = {3646-3662},
          doi = {10.1093/mnras/stz1815},
archivePrefix = {arXiv},
       eprint = {1810.05649},
 primaryClass = {astro-ph.CO},
       adsurl = {https://ui.adsabs.harvard.edu/abs/2019MNRAS.488.3646R}
}

@ARTICLE{2004ApJ...606..819M,
       author = {{Markevitch}, M. and {Gonzalez}, A.~H. and {Clowe}, D. and {Vikhlinin}, A. and {Forman}, W. and {Jones}, C. and {Murray}, S. and {Tucker}, W.},
        title = "{Direct Constraints on the Dark Matter Self-Interaction Cross Section from the Merging Galaxy Cluster 1E 0657-56}",
      journal = {\apj},
         year = 2004,
        month = may,
       volume = {606},
       number = {2},
        pages = {819-824},
          doi = {10.1086/383178},
archivePrefix = {arXiv},
       eprint = {astro-ph/0309303},
 primaryClass = {astro-ph},
       adsurl = {https://ui.adsabs.harvard.edu/abs/2004ApJ...606..819M}
}

@ARTICLE{2024NatAs.tmp....9H,
       author = {{HyeongHan}, Kim and {Jee}, M. James and {Cha}, Sangjun and {Cho}, Hyejeon},
        title = "{Weak-lensing detection of intracluster filaments in the Coma cluster}",
      journal = {Nature Astronomy},
         year = 2024,
        month = jan,
          doi = {10.1038/s41550-023-02164-w},
 primaryClass = {astro-ph.CO},
       adsurl = {https://ui.adsabs.harvard.edu/abs/2024NatAs.tmp....9H}
}

@ARTICLE{2015Natur.528..105E,
       author = {{Eckert}, Dominique and {Jauzac}, Mathilde and {Shan}, Huanyuan and {Kneib}, Jean-Paul and {Erben}, Thomas and {Israel}, Holger and {Jullo}, Eric and {Klein}, Matthias and {Massey}, Richard and {Richard}, Johan and {Tchernin}, C{\'e}line},
        title = "{Warm-hot baryons comprise 5-10 per cent of filaments in the cosmic web}",
      journal = {\nat},
         year = 2015,
        month = dec,
       volume = {528},
       number = {7580},
        pages = {105-107},
          doi = {10.1038/nature16058},
archivePrefix = {arXiv},
       eprint = {1512.00454},
 primaryClass = {astro-ph.CO},
       adsurl = {https://ui.adsabs.harvard.edu/abs/2015Natur.528..105E}
}

@ARTICLE{2020MNRAS.494.5473K,
       author = {{Kuchner}, Ulrike and {Arag{\'o}n-Salamanca}, Alfonso and {Pearce}, Frazer R. and {Gray}, Meghan E. and {Rost}, Agust{\'\i}n and {Mu}, Chunliang and {Welker}, Charlotte and {Cui}, Weiguang and {Haggar}, Roan and {Laigle}, Clotilde and {Knebe}, Alexander and {Kraljic}, Katarina and {Sarron}, Florian and {Yepes}, Gustavo},
        title = "{Mapping and characterization of cosmic filaments in galaxy cluster outskirts: strategies and forecasts for observations from simulations}",
      journal = {\mnras},
         year = 2020,
        month = jun,
       volume = {494},
       number = {4},
        pages = {5473-5491},
          doi = {10.1093/mnras/staa1083},
archivePrefix = {arXiv},
       eprint = {2004.08408},
 primaryClass = {astro-ph.GA},
       adsurl = {https://ui.adsabs.harvard.edu/abs/2020MNRAS.494.5473K}
}

@ARTICLE{2013ApJ...762L..30Z,
       author = {{Zitrin}, A. and {Meneghetti}, M. and {Umetsu}, K. and {Broadhurst}, T. and {Bartelmann}, M. and {Bouwens}, R. and {Bradley}, L. and {Carrasco}, M. and {Coe}, D. and {Ford}, H. and {Kelson}, D. and {Koekemoer}, A.~M. and {Medezinski}, E. and {Moustakas}, J. and {Moustakas}, L.~A. and {Nonino}, M. and {Postman}, M. and {Rosati}, P. and {Seidel}, G. and {Seitz}, S. and {Sendra}, I. and {Shu}, X. and {Vega}, J. and {Zheng}, W.},
        title = "{CLASH: The Enhanced Lensing Efficiency of the Highly Elongated Merging Cluster MACS J0416.1-2403}",
      journal = {\apjl},
         year = 2013,
        month = jan,
       volume = {762},
       number = {2},
          eid = {L30},
        pages = {L30},
          doi = {10.1088/2041-8205/762/2/L30},
archivePrefix = {arXiv},
       eprint = {1211.2797},
 primaryClass = {astro-ph.CO},
       adsurl = {https://ui.adsabs.harvard.edu/abs/2013ApJ...762L..30Z}
}

@ARTICLE{2016ApJ...819..114K,
       author = {{Kawamata}, Ryota and {Oguri}, Masamune and {Ishigaki}, Masafumi and {Shimasaku}, Kazuhiro and {Ouchi}, Masami},
        title = "{Precise Strong Lensing Mass Modeling of Four Hubble Frontier Field Clusters and a Sample of Magnified High-redshift Galaxies}",
      journal = {\apj},
         year = 2016,
        month = mar,
       volume = {819},
       number = {2},
          eid = {114},
        pages = {114},
          doi = {10.3847/0004-637X/819/2/114},
archivePrefix = {arXiv},
       eprint = {1510.06400},
 primaryClass = {astro-ph.GA},
       adsurl = {https://ui.adsabs.harvard.edu/abs/2016ApJ...819..114K}
}

@ARTICLE{2016MNRAS.461.2126S,
       author = {{Sebesta}, Kevin and {Williams}, Liliya L.~R. and {Mohammed}, Irshad and {Saha}, Prasenjit and {Liesenborgs}, Jori},
        title = "{Testing light-traces-mass in Hubble Frontier Fields Cluster MACS-J0416.1-2403}",
      journal = {\mnras},
         year = 2016,
        month = sep,
       volume = {461},
       number = {2},
        pages = {2126-2134},
          doi = {10.1093/mnras/stw1433},
archivePrefix = {arXiv},
       eprint = {1507.08960},
 primaryClass = {astro-ph.CO},
       adsurl = {https://ui.adsabs.harvard.edu/abs/2016MNRAS.461.2126S}
}

@ARTICLE{2021A&A...645A.140B,
       author = {{Bergamini}, P. and {Rosati}, P. and {Vanzella}, E. and {Caminha}, G.~B. and {Grillo}, C. and {Mercurio}, A. and {Meneghetti}, M. and {Angora}, G. and {Calura}, F. and {Nonino}, M. and {Tozzi}, P.},
        title = "{A new high-precision strong lensing model of the galaxy cluster MACS J0416.1-2403. Robust characterization of the cluster mass distribution from VLT/MUSE deep observations}",
      journal = {\aap},
         year = 2021,
        month = jan,
       volume = {645},
          eid = {A140},
        pages = {A140},
          doi = {10.1051/0004-6361/202039564},
archivePrefix = {arXiv},
       eprint = {2010.00027},
 primaryClass = {astro-ph.GA},
       adsurl = {https://ui.adsabs.harvard.edu/abs/2021A&A...645A.140B}
}

@ARTICLE{2022Natur.603..815W,
       author = {{Welch}, Brian and {Coe}, Dan and {Diego}, Jose M. and {Zitrin}, Adi and {Zackrisson}, Erik and {Dimauro}, Paola and {Jim{\'e}nez-Teja}, Yolanda and {Kelly}, Patrick and {Mahler}, Guillaume and {Oguri}, Masamune and {Timmes}, F.~X. and {Windhorst}, Rogier and {Florian}, Michael and {de Mink}, S.~E. and {Avila}, Roberto J. and {Anderson}, Jay and {Bradley}, Larry and {Sharon}, Keren and {Vikaeus}, Anton and {McCandliss}, Stephan and {Brada{\v{c}}}, Maru{\v{s}}a and {Rigby}, Jane and {Frye}, Brenda and {Toft}, Sune and {Strait}, Victoria and {Trenti}, Michele and {Sharma}, Soniya and {Andrade-Santos}, Felipe and {Broadhurst}, Tom},
        title = "{A highly magnified star at redshift 6.2}",
      journal = {\nat},
         year = 2022,
        month = mar,
       volume = {603},
       number = {7903},
        pages = {815-818},
          doi = {10.1038/s41586-022-04449-y},
archivePrefix = {arXiv},
       eprint = {2209.14866},
 primaryClass = {astro-ph.GA},
       adsurl = {https://ui.adsabs.harvard.edu/abs/2022Natur.603..815W}
}

@ARTICLE{2024Natur.628...57F,
       author = {{Furtak}, Lukas J. and {Labb{\'e}}, Ivo and {Zitrin}, Adi and {Greene}, Jenny E. and {Dayal}, Pratika and {Chemerynska}, Iryna and {Kokorev}, Vasily and {Miller}, Tim B. and {Goulding}, Andy D. and {de Graaff}, Anna and {Bezanson}, Rachel and {Brammer}, Gabriel B. and {Cutler}, Sam E. and {Leja}, Joel and {Pan}, Richard and {Price}, Sedona H. and {Wang}, Bingjie and {Weaver}, John R. and {Whitaker}, Katherine E. and {Atek}, Hakim and {Bogd{\'a}n}, {\'A}kos and {Charlot}, St{\'e}phane and {Curtis-Lake}, Emma and {van Dokkum}, Pieter and {Endsley}, Ryan and {Feldmann}, Robert and {Fudamoto}, Yoshinobu and {Fujimoto}, Seiji and {Glazebrook}, Karl and {Juneau}, St{\'e}phanie and {Marchesini}, Danilo and {Maseda}, Micheal V. and {Nelson}, Erica and {Oesch}, Pascal A. and {Plat}, Ad{\`e}le and {Setton}, David J. and {Stark}, Daniel P. and {Williams}, Christina C.},
        title = "{A high black-hole-to-host mass ratio in a lensed AGN in the early Universe}",
      journal = {\nat},
         year = 2024,
        month = apr,
       volume = {628},
       number = {8006},
        pages = {57-61},
          doi = {10.1038/s41586-024-07184-8},
archivePrefix = {arXiv},
       eprint = {2308.05735},
 primaryClass = {astro-ph.GA},
       adsurl = {https://ui.adsabs.harvard.edu/abs/2024Natur.628...57F}
}

@ARTICLE{2023ApJ...949L..34H,
       author = {{Hsiao}, Tiger Yu-Yang and {Coe}, Dan and {Abdurro'uf} and {Whitler}, Lily and {Jung}, Intae and {Khullar}, Gourav and {Meena}, Ashish Kumar and {Dayal}, Pratika and {Barrow}, Kirk S.~S. and {Santos-Olmsted}, Lillian and {Casselman}, Adam and {Vanzella}, Eros and {Nonino}, Mario and {Jim{\'e}nez-Teja}, Yolanda and {Oguri}, Masamune and {Stark}, Daniel P. and {Furtak}, Lukas J. and {Zitrin}, Adi and {Adamo}, Angela and {Brammer}, Gabriel and {Bradley}, Larry and {Diego}, Jose M. and {Zackrisson}, Erik and {Finkelstein}, Steven L. and {Windhorst}, Rogier A. and {Bhatawdekar}, Rachana and {Hutchison}, Taylor A. and {Broadhurst}, Tom and {Dimauro}, Paola and {Andrade-Santos}, Felipe and {Eldridge}, Jan J. and {Acebron}, Ana and {Avila}, Roberto J. and {Bayliss}, Matthew B. and {Ben{\'\i}tez}, Alex and {Binggeli}, Christian and {Bolan}, Patricia and {Brada{\v{c}}}, Maru{\v{s}}a and {Carnall}, Adam C. and {Conselice}, Christopher J. and {Donahue}, Megan and {Frye}, Brenda and {Fujimoto}, Seiji and {Henry}, Alaina and {James}, Bethan L. and {Kassin}, Susan A. and {Kewley}, Lisa and {Larson}, Rebecca L. and {Lauer}, Tod and {Law}, David and {Mahler}, Guillaume and {Mainali}, Ramesh and {McCandliss}, Stephan and {Nicholls}, David and {Pirzkal}, Norbert and {Postman}, Marc and {Rigby}, Jane R. and {Ryan}, Russell and {Senchyna}, Peter and {Sharon}, Keren and {Shimizu}, Ikko and {Strait}, Victoria and {Tang}, Mengtao and {Trenti}, Michele and {Vikaeus}, Anton and {Welch}, Brian},
        title = "{JWST Reveals a Possible z   11 Galaxy Merger in Triply Lensed MACS0647-JD}",
      journal = {\apjl},
         year = 2023,
        month = jun,
       volume = {949},
       number = {2},
          eid = {L34},
        pages = {L34},
          doi = {10.3847/2041-8213/acc94b},
archivePrefix = {arXiv},
       eprint = {2210.14123},
 primaryClass = {astro-ph.GA},
       adsurl = {https://ui.adsabs.harvard.edu/abs/2023ApJ...949L..34H}
}

@ARTICLE{2015Sci...347.1123K,
       author = {{Kelly}, Patrick L. and {Rodney}, Steven A. and {Treu}, Tommaso and {Foley}, Ryan J. and {Brammer}, Gabriel and {Schmidt}, Kasper B. and {Zitrin}, Adi and {Sonnenfeld}, Alessandro and {Strolger}, Louis-Gregory and {Graur}, Or and {Filippenko}, Alexei V. and {Jha}, Saurabh W. and {Riess}, Adam G. and {Bradac}, Marusa and {Weiner}, Benjamin J. and {Scolnic}, Daniel and {Malkan}, Matthew A. and {von der Linden}, Anja and {Trenti}, Michele and {Hjorth}, Jens and {Gavazzi}, Raphael and {Fontana}, Adriano and {Merten}, Julian C. and {McCully}, Curtis and {Jones}, Tucker and {Postman}, Marc and {Dressler}, Alan and {Patel}, Brandon and {Cenko}, S. Bradley and {Graham}, Melissa L. and {Tucker}, Bradley E.},
        title = "{Multiple images of a highly magnified supernova formed by an early-type cluster galaxy lens}",
      journal = {Science},
         year = 2015,
        month = mar,
       volume = {347},
       number = {6226},
        pages = {1123-1126},
          doi = {10.1126/science.aaa3350},
archivePrefix = {arXiv},
       eprint = {1411.6009},
 primaryClass = {astro-ph.CO},
       adsurl = {https://ui.adsabs.harvard.edu/abs/2015Sci...347.1123K}
}

@ARTICLE{2020MNRAS.498.1420W,
       author = {{Wong}, Kenneth C. and {Suyu}, Sherry H. and {Chen}, Geoff C. -F. and {Rusu}, Cristian E. and {Millon}, Martin and {Sluse}, Dominique and {Bonvin}, Vivien and {Fassnacht}, Christopher D. and {Taubenberger}, Stefan and {Auger}, Matthew W. and {Birrer}, Simon and {Chan}, James H.~H. and {Courbin}, Frederic and {Hilbert}, Stefan and {Tihhonova}, Olga and {Treu}, Tommaso and {Agnello}, Adriano and {Ding}, Xuheng and {Jee}, Inh and {Komatsu}, Eiichiro and {Shajib}, Anowar J. and {Sonnenfeld}, Alessandro and {Blandford}, Roger D. and {Koopmans}, L{\'e}on V.~E. and {Marshall}, Philip J. and {Meylan}, Georges},
        title = "{H0LiCOW - XIII. A 2.4 per cent measurement of H$_{0}$ from lensed quasars: 5.3{\ensuremath{\sigma}} tension between early- and late-Universe probes}",
      journal = {\mnras},
         year = 2020,
        month = oct,
       volume = {498},
       number = {1},
        pages = {1420-1439},
          doi = {10.1093/mnras/stz3094},
archivePrefix = {arXiv},
       eprint = {1907.04869},
 primaryClass = {astro-ph.CO},
       adsurl = {https://ui.adsabs.harvard.edu/abs/2020MNRAS.498.1420W}
}

@ARTICLE{2007arXiv0710.5636E,
       author = {{El{\'\i}asd{\'o}ttir}, {\'A}rd{\'\i}s and {Limousin}, Marceau and {Richard}, Johan and {Hjorth}, Jens and {Kneib}, Jean-Paul and {Natarajan}, Priya and {Pedersen}, Kristian and {Jullo}, Eric and {Paraficz}, Danuta},
        title = "{Where is the matter in the Merging Cluster Abell 2218?}",
      journal = {arXiv e-prints},
         year = 2007,
        month = oct,
          eid = {arXiv:0710.5636},
        pages = {arXiv:0710.5636},
          doi = {10.48550/arXiv.0710.5636},
archivePrefix = {arXiv},
       eprint = {0710.5636},
 primaryClass = {astro-ph},
       adsurl = {https://ui.adsabs.harvard.edu/abs/2007arXiv0710.5636E}
}

@ARTICLE{2015ApJ...800...84C,
       author = {{Coe}, Dan and {Bradley}, Larry and {Zitrin}, Adi},
        title = "{Frontier Fields: High-redshift Predictions and Early Results}",
      journal = {\apj},
         year = 2015,
        month = feb,
       volume = {800},
       number = {2},
          eid = {84},
        pages = {84},
          doi = {10.1088/0004-637X/800/2/84},
archivePrefix = {arXiv},
       eprint = {1405.0011},
 primaryClass = {astro-ph.GA},
       adsurl = {https://ui.adsabs.harvard.edu/abs/2015ApJ...800...84C}
}

\end{document}